\documentclass[12pt]{article}
\usepackage{hyperref}
\hypersetup{
    colorlinks,
    citecolor=blue,
    linkcolor = blue
}
\usepackage{graphicx, makecell, caption, multirow, adjustbox, hhline, supertabular, pdfpages, tikz, booktabs, color, color, colortbl, etoolbox, listings, enumerate, multirow, bigdelim, amsmath, amssymb, bbm, mathtools, ragged2e, amsthm, amsfonts, amstext, bm, longtable, rotating, dcolumn, lscape, threeparttable, geometry, csquotes, command, overpic, xparse, float, natbib, multirow, subcaption, enumitem, setspace, titlesec, threeparttable}
\usepackage{authblk}

\usepackage{algorithm}
\usepackage{algpseudocode}
\usepackage{float} 

\newtheorem{theorem}{Theorem}
\newtheorem{lemma}{Lemma}

\newtheorem{assumption}{Assumption}

\newcommand{\indep}{\perp \!\!\! \perp}

\newcommand{\tightdisplays}{%
    \setlength{\abovedisplayskip}{6pt plus 2pt minus 2pt}%
    \setlength{\belowdisplayskip}{6pt plus 2pt minus 2pt}%
    \setlength{\abovedisplayshortskip}{3pt plus 1pt minus 1pt}%
    \setlength{\belowdisplayshortskip}{3pt plus 1pt minus 1pt}%
}

\AddToHook{cmd/normalsize/after}{\tightdisplays}
\usepackage{bibunits}
\defaultbibliographystyle{apalike}
\defaultbibliography{bibliography}

\newcommand{\ipw}{\mathrm{I\text{-}IPW}}
\newcommand{\hj}{\mathrm{I\text{-}HJ}}

\newcommand{\Ybi}{\bar{Y}_i}                    
\newcommand{\Cbip}[1]{\bar{C}_i(#1)}            
\newcommand{\piclus}[1]{\pi_i(#1)}              
\newcommand{\wclus}[1]{w_i(#1)}                 
\newcommand{\DN}[1]{D_N(#1)}                    

\DeclareMathOperator{\Reg}{Reg}                 

\newcommand{\Ybar}[1]{\bar{Y}(#1)}
\newcommand{\Cbar}[1]{\bar{C}(#1)}

\newcommand{\Yhat}[1]{\widehat{\bar{Y}}_{#1}}
\newcommand{\Chat}{\widehat{\bar{C}}}
\newcommand{\Vhat}[1]{\widehat{V}_{#1}}
\newcommand{\phat}[1]{\widehat{p}_{#1}}
\newcommand{\sdhat}[1]{\widehat{\sigma}_{#1}}
\newcommand{\Bdev}[1]{B_{#1}}
\newcommand{\alphahat}[1]{\widehat{\alpha}_{N,#1}}
\newcommand{\psiipw}{\psi_i^{\ipw}}
\newcommand{\phihj}{\phi_i^{\hj}}

\hypersetup{
    colorlinks,
    citecolor=blue,
    linkcolor=blue,
    urlcolor=blue
}

\DeclareMathOperator{\Varop}{Var}

\title{Constrained Stochastic Policy Targeting under Heterogeneous Partial Interference}

\author[1]{Elena Dal Torrione}
\author[2]{Chan Park}
\author[1]{Laura Forastiere}

\affil[1]{\small Department of Biostatistics, Yale University}
\affil[2]{Department of Statistics, University of Illinois Urbana-Champaign}

\date{\small This version: November 13, 2026\\[2pt]
\href{https://edaltorrione.github.io/files/Constrained_Stochastic_Policy_Targeting.pdf}{Click here for the latest version}}

\titlespacing*{\section}{0pt}{10pt}{4pt}
\titlespacing*{\subsection}{0pt}{8pt}{3pt}
\titlespacing*{\subsubsection}{0pt}{6pt}{2pt}

\begin{document}
\begin{onehalfspace}
\maketitle
\end{onehalfspace}

\tightdisplays

\setcounter{page}{1}

\vspace{-1em}

\begin{abstract}
Under heterogeneous interference, treating individuals with particular characteristics affects the population average outcome through their heterogeneous direct response to treatment and indirect effect on others, creating opportunities for efficient policy design. For instance, policymakers may minimize resource use by vaccinating individuals identified as superspreaders to achieve a target reduction in disease incidence.
In such setting, we first study the problem of learning individualized stochastic treatment policies that minimize population average treatment probability subject to a prespecified outcome target. 
We propose an estimation approach based on a smooth surrogate for constraint violations and interference-aware inverse-probability-weighting or self-normalized estimators of the population average outcome, and establish finite-sample regret guarantees for both estimators. We also consider the complementary problem of maximizing the population average outcome subject to a constraint on average treatment probability. We apply the proposed methodology to a Colombian conditional cash transfer program aimed at improving school attendance. The application demonstrates that accounting for interference and individualizing treatment probabilities can reduce resource use relative to policies that ignore interference or impose homogeneous treatment probabilities.
\end{abstract}


\section{Introduction}

Many interventions are implemented in settings in which individuals interact, so that the
treatment assigned to one individual may affect not only that individual's outcome but
also the outcomes of others. Such phenomenon, known as interference, arises naturally within households,
schools, villages, and other socially or physically connected groups \citep{cox1958}.
Accounting for these dependencies
is particularly important for optimal policy learning, 
aimed at learning treatment rules that optimize a population-level objective
\citep{athey2021policy, kitagawa2018ewm, zhao2012owl,QianMurphy2011}. Without interference, individualized treatment rules
exploit heterogeneity in individuals' responses to their own treatment. Under interference,
heterogeneity can also arise in the effects that one individual's treatment has on others, which may depend on their characteristics and those of connected individuals and on the interaction structure in the population \citep{bargaglistoffi2025,dean2025,viviano2024}. For instance, in the Colombian conditional cash
transfer program \textit{Subsidios Condicionados a la Asistencia Escolar}, prior work documents both heterogeneity in treatment response by gender and socioeconomic status and heterogeneous adverse spillovers on untreated siblings, with larger
attendance reductions among female siblings of treated students \citep{osorio,ferreira}.

In such settings, accounting for interference is consequential for policy design for two reasons. First, a policy-learning procedure that disregards interference can lead to biased estimates of relevant policy quantities and select an allocation that is suboptimal for the population objective
\citep{viviano2024,zhang2024}. Second, heterogeneity in direct and spillover effects can
be leveraged to design efficient policy targeting when resources are scarce. If spillovers
are beneficial, prioritizing individuals whose treatment generates larger gains for others
can attain a prespecified
outcome using fewer resources or increase population welfare for a given treatment budget. If spillovers are harmful, an interference-aware policy can instead avoid allocations that generate negative indirect effects on others' outcomes.

Most of the policy-learning literature considers the problem of assigning treatment to
maximize population welfare, possibly subject to a constraint on available resources
\citep{KosorokMoodie2015,kitagawa2018ewm,athey2021policy}.
A complementary problem arises when the policymaker instead faces a prespecified outcome
target and seeks the least resource-intensive policy that attains it \citep{BatthacharyaDupas2012}. In this paper, we consider both problems under partial interference, where interference may occur within given clusters but not across clusters. Specifically, 
let $p\in\mathcal P$ denote a covariate-dependent treatment policy, $\overline{Y}(p)$ the population average outcome induced by the policy, and
$\overline{C}(p)$ its population average treatment intensity. We study
$\max_{p\in\mathcal{P}} \overline{Y}(p)$ subject to $\overline{C}(p) \leq B$, and 
$\min_{p\in\mathcal{P}} \overline{C}(p)$ subject to $\overline{Y}(p) \geq T$.
The first formulation, which we refer to as the \qq resource-constrained" problem, is a version of the familiar welfare-maximization problem under a resource budget. In contrast, 
the second formulation, referred to as
the \qq outcome-target" problem, selects the least intensive treatment policy that attains a prespecified level of population outcome. 
Although understudied, the outcome-target problem is particularly relevant when policymakers face externally specified performance goals, such as the UN Sustainable Development Goals and WHO targets for malaria reduction \citep{unitednations2016transforming,who2021malaria}. 
For instance, in the Colombian cash transfer program, a policymaker may seek the least intensive cash-transfer policy that attains a specified level of school attendance.
The logic is analogous to sample-size planning, where a desired level of statistical power is specified first and the required sample size is then determined. In this paper, our primary focus is the outcome-target problem; we then extend the analysis to the resource-constrained problem.
Under partial interference, policy evaluation involves high-dimensional cluster-level treatment assignments. This poses challenges for deterministic policies, since inverse-probability-weighted (IPW) evaluation 
requires exact matches between observed and policy-prescribed assignments, which can be rare in finite samples \citep{zhang2024}. We therefore consider individualized stochastic treatment policies that assign treatment probabilities based on individual- and cluster-level covariates. Such policies allow multiple observed treatment configurations to contribute to estimation \citep{kennedy2019, swaminathan2015crm} and, for smooth parametric classes, allow for efficient gradient-based optimization \citep{luedtke2016, fangwangwang2023}. Moreover, under interference, stochastic policies can be optimal and strictly outperform deterministic allocations \citep{DalTorrione_etal_2026}\footnote{
\citet[Chapter 1.3]{bickel2015mathematical} establish that randomized decision rules strictly outperform deterministic ones when the risk set of the deterministic rules is non-convex. Similarly, under interference, stochastic policies outperform deterministic ones when the outcome space is non-convex due to interactions. For instance, consider a cluster of units with identical covariates where a mixed treatment allocation yields a greater cluster-average outcome than uniform allocations. Because any deterministic covariate-based rule is constrained to treat either all units or no units within the cluster, deterministic rules yield smaller average outcomes, whereas an independent stochastic policy achieves a greater expected outcome by convexifying the action space.
}.

The estimated constraint creates a challenge absent from standard unconstrained policy
learning, since sampling error can make an infeasible policy 
appear feasible \citep{sun2026}. 
To address this problem, we develop an
empirical policy estimator that tightens the estimated constraint through a margin $\epsilon$ and
uses a smooth surrogate for constraint violation. We establish non-asymptotic regret
guarantees for the resulting IPW estimator, where the regret bound balances the surrogate approximation error with the objective and constraint estimation error, yielding
a regret of order $N^{-1/4}$.
Under additional conditions, we also obtain the sharper rate $O\{\sqrt{\log(N)/N}\}$. 

Under interference, IPW estimators may suffer from extreme weights leading to high variance due to the reweighting of joint treatment configurations at the cluster level. Furthermore, because this variance profile can differ substantially across policies, enforcing a uniform margin can be excessively conservative.
We address both issues in two complementary ways. First, alongside the interference-aware IPW estimator, we consider its self-normalized, or H\'ajek, analogue \citep{hajek1971,TchegenVanderweele2012,papadogeorgou19}, which can stabilize cluster-level weights. Second, drawing on empirical Bernstein arguments \citep{maurerpontil,swaminathan2015crm}, we develop policy-specific, variance-adaptive constraint margins for both IPW and Hájek estimation, which impose stricter feasibility requirements on policies with more variable outcome estimates and less conservative requirements on those estimated more precisely.
We establish finite-sample regret guarantees for the Hájek estimator, which attain the same rates as the IPW estimator, and for the variance-adaptive procedures under both IPW and Hájek estimation.
%
%
To our knowledge, these are the first uniform empirical Bernstein bounds for interference-aware IPW and Hájek policy evaluation, and may be of independent statistical interest.
%

Finally, we study the finite-sample performance of the proposed methods in simulations and apply
them to a Colombian conditional cash transfer experiment with potential within-household interference.

\paragraph{Related Literature}

Our work bridges the two distinct strands of literature concerning causal inference under interference and optimal
treatment policy learning. An extensive body of work on interference develops identification
and estimation methods for direct, spillover, and overall causal effects
\citep{HudgensHalloran2008,TchegenVanderweele2012,aronowsamii17,forastiere,leung2020treatment,DalTorrione_etal_RDD,Park2022}, while also exploring effect heterogeneity and covariate-dependent allocation strategies
\citep{bargaglistoffi2025,dean2025,papadogeorgou19}. The policy-learning literature
\citep{athey2021policy, kitagawa2018ewm, zhao2012owl, QianMurphy2011} operates predominantly under the assumption of  no interference. While an emerging domain intersects these fields by studying  treatment allocation and policy learning under interference
\citep{kitagawa23,ananth24,viviano2024,vivianorudder2024,zhang2024}, most existing
approaches focus on welfare maximization and deterministic treatment rules, often
imposing restrictive structural assumptions on the interference mechanism to maintain tractability. We instead study constrained policy learning with stochastic policies under partial interference, allowing potential outcomes to depend arbitrarily on the treatment assignments of all units within a cluster.

Our outcome-target formulation is most closely related to \citet{park24}, who introduce
minimum resource threshold policies determining the minimum proportion of treated units
required for a cluster to attain a cluster-level outcome target. We instead aim to learn individualized treatment
probabilities as functions of observed characteristics, while imposing the
constraint on the population average outcome. In no-interference settings,
\citet{BatthacharyaDupas2012} characterize budget-constrained allocations across
unrestricted policy classes using individual effect-to-cost ratios; by contrast, we learn the optimal policy a prespecified class while explicitly allowing for spillover effects. More recently, \citet{sun2026} studies welfare maximization under an estimated budget constraint for deterministic eligibility policies, deriving asymptotic results on welfare efficiency and feasibility. We instead consider stochastic policies under interference, minimizing treatment intensity subject to an outcome target, and derive finite-sample regret guarantees for interference-aware estimators.
Finally, our work relates to self-normalized off-policy estimation, which has received considerable attention in the machine-learning literature \citep[e.g.,][]{swaminathan2015selfnormalized, kuzborskij2021confident}. We contribute uniform finite-sample empirical Bernstein bounds for interference-aware IPW and H\'ajek estimators over potentially infinite policy classes, together with their implications for constrained policy regret.

\paragraph{Organization of the paper.}
Section~\ref{sec:setup} introduces the partial-interference framework, stochastic policy
class, and outcome-target optimization problem. Section~\ref{sec:estimation} develops our proposed policy estimators.
Section~\ref{sec:theory} establishes the regret guarantees, and
Section~\ref{sec:extension} develops the variance-adaptive constraint margins.
Section~\ref{sec:dualproblem} discusses the complementary resource-constrained problem of maximizing welfare subject to a constraint on the average treatment probability. Sections~\ref{sec:simulation} and~\ref{sec:empirical} present the simulation study and the empirical application, and Section~\ref{sec:conclusion} concludes. Appendix~\ref{appendix:dualproblem} provides the detailed analysis of the resource-constrained problem, Appendix~\ref{appendix:additional} reports additional tables for the empirical application, and Appendix~\ref{appendix:proof} contains the proofs.

\section{Framework}\label{sec:setup}

\subsection{Setup and Notation}
Let $N$ be the total number of clusters, with each cluster indexed by $i=1,\ldots,N$. For each cluster $i$, $n_i$ is the number of units and units are indexed by $j=1,\ldots,n_i$, such that the subscript $ij$ denotes unit $j$ in cluster $i$. For each unit $j$ in cluster $i$ we observe the treatment assignment $A_{i j} \in\{0,1\}$, the outcome $Y_{ij} \in \mathbbm{R}$ and $K$ individual covariates $\bm X_{ij} \in \mathcal{X}$, with $\mathcal{X}\subseteq \mathbbm{R}^{K}$. In addition, each cluster carries $L$ cluster-level covariates $\bm V_i \in \mathcal V \subseteq \mathbbm{R}^L$. 
Without loss of generality, we assume that a larger value of $Y_{ij}$ is preferred.
Finally, for each cluster $i$, we use $\boldsymbol{A}_i \in \mathcal{A}\left(n_i\right)=\{0,1\}^{n_i}$, $\bm X_i=\left(\boldsymbol{X}_{i 1} \; \boldsymbol{X}_{i 2} \; \cdots\; \boldsymbol{X}_{i{n_i}}\right)^{\top}$, and  $\boldsymbol{Y}_i=\left(Y_{i 1} \; Y_{i 2}\; \cdots\; Y_{i n_i}\right)$ to denote the vector of treatment, the $n_i \times K$ matrix of covariates, and the vector of observed outcomes, respectively. 
Let us denote by $\bm O_i = \{\bm Y_i, \bm A_i, \bm X_i, \bm V_i\}$ the observed data in the cluster. We assume that $\{\bm O_i\}_{i = 1}^{N}$ is an $i.i.d.$ sample drawn from a superpopulation of clusters characterized by (unknown) distribution $G_0$.
Throughout, lower case letters indicate realizations of the corresponding random variables.

In the presence of interference, the potential outcome of an individual may depend on the treatment of other individuals. In full generality, let $Y_{i j}\left(\boldsymbol{a}_1, \boldsymbol{a}_2, \ldots, \boldsymbol{a}_N\right)$ denote the potential outcome for unit $j$ in cluster $i$ when the treatment of units across all clusters is equal to $(\boldsymbol{a}_1, \boldsymbol{a}_2, \ldots, \boldsymbol{a}_N)$ for $\boldsymbol{a}_i \in \mathcal{A}\left(n_i\right)$.
Throughout the paper, we make a partial (clustered) interference assumption, which restricts interference to occur only within clusters but not across clusters \citep{sobel, HudgensHalloran2008}. Formally,

\begin{assumption}[Partial Interference]\label{ass:partial}
For all $i=1,\ldots,N$ and $j=1,\ldots,n_i$, and for all vector of cluster treatment assignments $(\bm a_1, \ldots, \bm a_N)$ and $(\bm a_1', \ldots, \bm a_N')$ such that $\bm a_i = \bm a_i^{\prime}$, we have that $Y_{i j}\left(\bm a_1, \ldots, \bm a_N\right)=Y_{i j}\left(\bm a_1', \ldots, \bm a_N'\right)$.
\end{assumption}

Under this assumption, for each unit $j$ in cluster $i$, let $Y_{ij}(\bm a_i)$ denote the potential outcome under cluster treatment $\bm a_i$, representing the outcome that would be observed for unit $j$ had the treatment vector in cluster $i$ been set to $\bm a_i$.

\subsection{Individualized Stochastic Treatment Policy}

We consider stochastic, covariate-dependent treatment allocations where each individual is treated independently with a probability determined by their individual and cluster-level covariates. Specifically, we define an individualized stochastic treatment policy as a function from the individual and cluster-level space to the interval $[0,1]$, $p : \mathcal{X} \times\mathcal{V} \mapsto [0,1]$. Hence, $p(\bm X_{ij}, \bm V_{i}) = \Prr_p(A_{ij} = 1 \mid \bm X_{ij}, \bm V_i)$ specifies the conditional probability of treatment for an individual with characteristics $(\bm X_{ij}, \bm V_{i})$. 
Since the policy imposes independent treatment assignment conditional on covariates, the probability of a treatment vector $\boldsymbol{a}_i \in \mathcal{A}\left(n_i\right)$ in cluster $i$ under policy $p$ is
\begin{equation}\label{eq:clusterprob}
\piclus{\boldsymbol{a}_i \mid \boldsymbol{X}_i, \boldsymbol{V}_i;p}
\;:=\;
\mathbb{P}_p\left(\boldsymbol{A}_i=\boldsymbol{a}_i \mid \boldsymbol{X}_i, \boldsymbol{V}_i\right)
=\prod_{j=1}^{n_i} p\left(\boldsymbol{X}_{i j}, \boldsymbol{V}_i\right)^{a_{i j}}\left(1-p\left(\boldsymbol{X}_{i j}, \boldsymbol{V}_i\right)\right)^{1-a_{ij}} .
\end{equation}

We aim to choose an optimal policy $p$ within a prespecified class of functions $\mathcal{P}$. 
We assume that $\mathcal P$ contains at least one constant policy. Possible function classes in the present work include the class of linear probability functions $\mathcal{P}_{\Gamma,\mathrm{lin}}=\{(\boldsymbol{x}, \boldsymbol{v}) \mapsto\langle\gamma,(\boldsymbol{x}, \boldsymbol{v})\rangle: \gamma \in \Gamma\}$ for some parameter space $\Gamma$, and the class of logistic probability functions $\mathcal{P}_{\Gamma,\mathrm{logit}}=\left\{(\boldsymbol{x}, \boldsymbol{v}) \mapsto\sigma(\langle\gamma,(\boldsymbol{x}, \boldsymbol{v})\rangle): \gamma \in \Gamma\right\}$ with $\sigma(u) = (1+\exp (-u))^{-1}$ \citep[e.g.][]{swaminathan2015crm, papadogeorgou19, dean2025}.

\subsection{Optimization Problem}\label{sec:optprobl}

Adapting the definition in \citet{HudgensHalloran2008} and \citet{TchegenVanderweele2012}, we define the population average potential outcome under policy $p$ as 
\[
\Ybar{p} = \E_{G_0}\left\{\frac{1}{n_i}\sum_{j=1}^{n_i} \sum_{\bm a_i \in \mathcal{A}\left(n_i\right)} Y_{i j}\left(\boldsymbol{a}_i\right) \piclus{\bm a_i \mid \bm X_i, \bm V_i;p}\right\}
\]
Similarly we define the population average treatment probability as 
\[
\Cbar{p} = \E_{G_0}\left\{\frac{1}{n_i}\sum_{j=1}^{n_i} p(\bm X_{ij},\bm V_i)\right\}
\]
which represents a measure of resource use of policy $p$. 
%
Our problem consists in finding a policy $p^\star$ in a class of policies $\mathcal{P}$ which minimizes the population treatment propensity while achieving a prespecified outcome level:
%
\begin{equation}\label{eq:minproblem}
p^\star \in \arg\min_{p \in \mathcal{P}} \Cbar{p}
\qquad
\text{s.t.} \quad \Ybar{p} \geq T .
\end{equation}
%
Problem \eqref{eq:minproblem} can be conveniently rewritten in an equivalent unconstrained form as
\begin{align}\label{eq:originalproblem}
    p^\star \in \arg\min_{p \in \mathcal{P}} R_{G_0}(p),
    \quad \text{ where } \quad 
    R_{G_0} (p) := \Cbar{p} + C_0 \cdot \bm{1}\big\{\Ybar{p} < T\big\} . 
\end{align}
%
where $C_0$ is a constant larger than $1$. We refer to $R_{G_0}(p)$ as the \qq risk" of policy $p$. This risk definition incorporates both our preference for minimizing resource use and the penalty associated with violating the outcome constraint. 

The population average outcome is not identified from the observed data without further assumptions. Let $f(\bm A_i = \bm a_i \mid \bm X_i, \bm V_i)$ denote the known cluster treatment propensity score under $G_0$. For causal identification, we impose the following assumptions, which extend consistency, conditional unconfoundedness, and overlap to partial interference settings \citep[e.g.][]{TchegenVanderweele2012,Park2022}.

\begin{assumption}\label{ass:identification}
For all $i=1,\ldots,N$ and $j=1,\ldots,n_i$ the following holds:
    \begin{enumerate}[
    label=\alph*),
    itemsep=0pt,
    parsep=0pt,
    topsep=2pt,
    partopsep=0pt
]
    \item Consistency: $\bm Y_i = \bm Y_i(\bm{A}_i)$.
    \item[b)] Unconfoundedness: $\bm Y_i(\bm{a}_i) \indep \bm{A}_i \, | \, \bm{X}_i, \bm{V}_i$ for all $\bm{a}_i \in \mathcal{A}(n_i)$.
    \item[c)] Positivity: for some $\eta >0$, $ f(\bm{A}_i = \bm{a}_i \mid \bm{X}_i, \bm{V}_i) > \eta$ for all $\bm{a}_i \in \mathcal{A}(n_i)$ such that $\piclus{\bm a_i \mid \bm X_i, \bm V_i;p} > 0$ for some $p \in \mathcal{P}$.\label{ass:overlap}
\end{enumerate}
\end{assumption}
\noindent 
Under Assumptions \ref{ass:partial} and \ref{ass:identification}, Problem \eqref{eq:originalproblem} can be rewritten in terms of observed data as
\begin{equation}\label{eq:probid}
p^\star \in \arg\min_{p \in \mathcal{P}} \;
\left[\Cbar{p} + C_0 \cdot \bm{1}\left\{\E_{G_0} \left\{\frac{1}{n_i}\sum_{j = 1}^{n_i} Y_{ij} w_i(p)
\right\}  < T\right\}\right], \quad  \wclus{p} = \frac{\piclus{\bm A_i \mid \bm X_i, \bm V_i;p}}{f(\bm A_i \mid \bm X_i,\bm V_i)} 
\end{equation}
where the weights $w_i(p)$
are the ratio of the probability of the observed cluster assignment under $p$ to its probability under the data-generating assignment mechanism. 

Since the risk depends on the unknown population distribution $G_0$, our goal is to estimate a policy $\hat p$ from the data. For this purpose, we use an empirical analog of the risk in the spirit of the Empirical Welfare Maximization framework (EWM), which maximizes a sample analog of the population welfare over a policy class
\citep{kitagawa2018ewm}. 
As in that literature, a learned policy can be evaluated by the population risk it attains relative to the best attainable risk in \(\mathcal P\), that is, by its regret. Accordingly, we define
\[
R_{G_0}^\star
:=
\inf_{p\in\mathcal P}R_{G_0}(p),
\]
and we evaluate an estimated policy \(\widehat p\) through its expected regret, defined as
\begin{equation}\label{eq:regret}
\Reg(\hat p) = \E_{\mathcal{O}}[R_{G_0}(\widehat p)]-R_{G_0}^\star.
\end{equation}
where the expectation is taken over the distribution of the observed data $(\bm O_1,\ldots, \bm O_N)$, sampled i.i.d.\ from the population distribution $G_0$. Hence, given this performance criterion, in the following analysis we do not require $p^\star$ to be unique or $\hat{p}$ to converge to a specific minimizer of $R_{G_0}$ over $\mathcal{P}$.
For \(t\in\mathbbm R\), we further denote by
\[
\bar C^\star(t)
:=
\inf_{p\in\mathcal P}\left\{\Cbar{p}:\ \Ybar{p}\geq t\right\}
\]
the minimal resource use required to attain outcome level \(t\) within \(\mathcal P\). Since \(\Cbar{p}\leq 1<C_0\), whenever some \(p\in\mathcal P\) satisfies \(\Ybar{p}\geq T\) we have \(R_{G_0}^\star=\bar C^\star(T)\).




\section{Estimation and Optimization}\label{sec:estimation}

To estimate the optimal policy, one could replace the population quantities in \eqref{eq:originalproblem} with their empirical analogues and minimize the resulting criterion. However, the indicator for constraint violation makes the objective discontinuous in the policy, complicating both theoretical analysis and numerical optimization. 
To this end, we replace the indicator with a smooth surrogate.
%
For $\epsilon>0$, define 
\begin{equation}\label{eq:smoothhinge}
h_\epsilon(u;T)
=
\begin{cases}
1,
& u \leq T,\\[4pt]
1-\dfrac{2}{\epsilon^2}(u-T)^2,
& T < u < T+\dfrac{\epsilon}{2},\\[8pt]
\dfrac{2}{\epsilon^2}(T+\epsilon-u)^2,
& T+\dfrac{\epsilon}{2}\leq u<T+\epsilon,\\[8pt]
0,
& u \geq T+\epsilon.
\end{cases}
\end{equation}
The function $h_\epsilon(u;T)$ upper bounds the constraint-violation indicator $\bm{1}(u<T)$, equals one for $u\leq T$, and vanishes for $u\geq T+\epsilon$. The parameter $\epsilon$ determines the width of the buffer region $[T,T+\epsilon]$, over which the penalty decreases smoothly from one to zero. Moreover,
$h_\epsilon(u;T)\to\bm{1}(u<T)$ pointwise for every $u\neq T$ as $\epsilon\to0$. Unlike the truncated linear hinge, $h_\epsilon(\cdot;T)$ is continuously differentiable and $2/\epsilon$-Lipschitz.
Using this surrogate, we consider
\begin{equation}\label{eq:surrogateprobl}
\min_{p\in\mathcal P}\;
V_{\epsilon,G_0}(p),
\qquad
V_{\epsilon,G_0}(p) \;=\; \Cbar{p} + C_0\, h_\epsilon\big(\Ybar{p};T\big),
\end{equation}
and refer to $V_{\epsilon,G_0}(p)$ as the surrogate risk. We denote by $V_{\epsilon,G_0}^\star:=\inf_{p\in\mathcal P}V_{\epsilon,G_0}(p)$ the optimal surrogate risk and by $p_\epsilon^\star\in\arg\min_{p\in\mathcal P}V_{\epsilon,G_0}(p)$ a surrogate-optimal policy. The surrogate risk
assigns penalty $C_0$ to infeasible policies with $\Ybar{p}<T$, zero penalty to policies satisfying the constraint with a margin, that is $\Ybar{p}\geq T+\epsilon$, and an intermediate penalty to policies with $\Ybar{p}\in[T,T+\epsilon]$. 
Since $h_\epsilon(u ; T) \geq \bm{1}(u<T)$, the surrogate constraint provides a conservative approximation to the original constraint, so $V_{\epsilon, G_0}(p) \geq R_{G_0}(p)$ for every $p \in \mathcal{P}$; in particular, at a surrogate-optimal solution, $V_{\epsilon, G_0}\left(p_\epsilon^{\star}\right) \geq R_{G_0}\left(p_\epsilon^{\star}\right)$. Whenever there exists a policy $p_\epsilon^0\in\mathcal P$ such that $\Ybar{p_\epsilon^0}\geq T+\epsilon$, we have $V_{\epsilon,G_0}(p_\epsilon^0)=\Cbar{p_\epsilon^0}\leq 1<C_0$, whereas every infeasible policy has surrogate risk at least $C_0$. Hence an optimal $p_\epsilon^{\star}$ is feasible for the original constraint, $\bar{Y}\left(p_\epsilon^{\star}\right) \geq T$, and under this condition the following also holds:
$$
V_{\epsilon,G_0}^\star=V_{\epsilon, G_0}\left(p_\epsilon^{\star}\right) \geq R_{G_0}\left(p_\epsilon^{\star}\right)=\bar{C}\left(p_\epsilon^{\star}\right) \geq R_{G_0}^\star=\bar{C}^\star(T),
$$
so that $V_{\epsilon,G_0}^\star \geq R_{G_0}^\star$ and $\bar{C}\left(p_\epsilon^{\star}\right) \geq \bar{C}^\star(T)$: the surrogate-optimal policy uses, on average, at least as much resource as an optimal policy for the original problem.


\subsection{Empirical Objective}\label{sec:constraintestimators}

The average treatment probability 
$\bar{C}(p)$ is estimated by its sample average, given by
\[
\Chat(p) = \frac{1}{N}\sum_{i=1}^N \Cbip{p}, \qquad \Cbip{p} = \frac{1}{n_i}\sum_{j=1}^{n_i} p(\bm X_{ij},\bm V_i).
\]
For the outcome constraint, we consider two estimators.
We first define the interference-aware IPW estimator, $\Yhat{\ipw}(p)$, as follows: 
\begin{equation}\label{eq:ipwscore}
\Yhat{\ipw}(p) \;=\; \frac{1}{N}\sum_{i=1}^N \psiipw(p) , \quad \text{ with  } \psiipw(p) \;=\; \frac{1}{n_i}\sum_{j = 1}^{n_i} Y_{ij} \frac{\piclus{\bm A_i \mid \bm X_i, \bm V_i;p}}{f(\bm A_i \mid \bm X_i,\bm V_i)}  ,
\end{equation}
The IPW estimator weights the average observed outcomes  within clusters by the ratio of the probability of the observed cluster-level observed treatment assignment under the target policy to that under the observed treatment mechanism, given by the $f(\bm A_i \mid \bm X_i, \bm V_i)$, which, throughout, we assume to be known.
By \eqref{eq:probid}, $\E_{G_0}\{\psiipw(p)\} = \Ybar{p}$, so $\Yhat{\ipw}(p)$ is unbiased for the population average outcome under policy $p$ 
under Assumptions \ref{ass:partial} and \ref{ass:identification} \citep{TchegenVanderweele2012, papadogeorgou19}. 

The IPW estimator in \eqref{eq:ipwscore} can be highly variable  especially when $n_i$ is large. As a potentially more stable alternative,  we therefore also consider the self-normalized, or H\'ajek, estimator, $\Yhat{\hj}(p)$, defined as follows 
\begin{equation}\label{eq:denominator}
\Yhat{\hj}(p) = \frac{N^{-1}\sum_{i=1}^N \Ybi\,\wclus{p}}{\DN{p}}, \quad \text{ with } \DN{p} = \frac{1}{N}\sum_{i=1}^N \wclus{p},
\end{equation}
where $\Ybi=n_i^{-1}\sum_{j=1}^{n_i}Y_{ij}$ is the observed cluster-average outcome.
Similar cluster-level weights have been used previously in the interference
literature \citep{papadogeorgou19,dean2025}. Although the H\'ajek estimator is generally biased in finite samples, it is consistent for the population average outcome, and the weight standardization can yield efficiency gains relative to the unstandardized IPW estimator \citep{hajek1971,liu2016ipw}.
%
%
We propose the following optimal policy estimators:
\[
\phat{\ipw}\in\arg\min_{p\in\mathcal P}\Vhat{\epsilon,\ipw}(p),
\qquad
\phat{\hj}\in\arg\min_{p\in\mathcal P}\Vhat{\epsilon,\hj}(p).
\]
with
\begin{equation}\label{eq:empiricalsurrogate}
\Vhat{\epsilon,\ipw}(p)
=
\Chat(p)
+
C_0\, h_\epsilon\big(\Yhat{\ipw}(p);T\big),
\qquad
\Vhat{\epsilon,\hj}(p)
=
\Chat(p)
+
C_0\, h_\epsilon\big(\Yhat{\hj}(p);T\big),
\end{equation}
Both estimators $\phat{\ipw}$ and $\phat{\hj}$ minimize the empirical analogue of $V_{\epsilon,G_0}(p)$, denoted by $\Vhat{\epsilon,\ipw}(p)$ and $\Vhat{\epsilon,\hj}(p)$, obtained by replacing $\bar C(p)$ with the sample average treatment propensity $\Chat(p)$, and $\bar Y(p)$ with the corresponding population average outcome estimator, $\Yhat{\ipw}(p)$ and $\Yhat{\hj}(p)$. The smooth surrogate, together with the use of stochastic policies, permits efficient gradient-based optimization. In Sections~\ref{sec:simulation} and~\ref{sec:empirical}, we minimize \eqref{eq:empiricalsurrogate} using the L-BFGS-B algorithm \citep{liu1989}, as implemented in the R \texttt{optim} function.

\section{Regret Analysis}\label{sec:theory}

To establish the theoretical properties of our policy estimators, we make the following assumptions.

\begin{assumption}\label{ass:regularity}
The following holds for all $i=1,\ldots,N$ and $j=1,\ldots,n_i$: 
\begin{enumerate}[
    label=\alph*),
    itemsep=0pt,
    parsep=0pt,
    topsep=2pt,
    partopsep=0pt
]    
    \item Bounded outcomes: there exists a constant $M$ such that $|Y_{ij}(\bm{a}_i)| < M$ for all $\bm{a}_i$.\label{ass:boundedoutcome}
    \item Bounded cluster size: there exists a constant $n_{max}$ such that $n_i < n_{max}$.\label{ass:boundedclus}
\end{enumerate}
\end{assumption}

\noindent Assumption \ref{ass:regularity}.\ref{ass:boundedoutcome} assumes bounded outcomes and is standard in the literature \citep{kitagawa2018ewm,zhang2024}. Assumption \ref{ass:regularity}.\ref{ass:boundedclus} imposes bounded cluster size $n_i$, ensuring that the cluster size is small relative to the number of clusters $N$ under the asymptotic regime where $N \rightarrow \infty$. 

\subsection{Regret for the IPW Estimator}
\label{sec:ipw-regret}

Our goal is to bound the regret of $\phat{\ipw}$, $\Reg(\phat{\ipw})$, as defined in Eq.~\eqref{eq:regret}.
\begin{theorem}\label{th:regret}
Under Assumptions \ref{ass:partial}--\ref{ass:regularity},
\begin{equation*}
   \Reg(\phat{\ipw})
   \leq  
   \underbrace{ V_{\epsilon,G_0}^\star - R_{G_0}^\star
   }_{(a)} + \underbrace{ 8n_{max} \mathcal R_N(\mathcal{P})}_{(b)} + 
   \underbrace{ 8\sqrt{2}\,n_{max}^{3/2}\frac{M}{\eta}\frac{C_0}{\epsilon} \mathcal R_N(\mathcal{P})+\frac{4C_0M}{\eta\epsilon}\sqrt{\frac{2\log2}{N}} }_{(c)} \ .
\end{equation*}
where 
$\mathcal R_N(\mathcal{P})$ denotes the expected Rademacher complexity of the function class $\mathcal{P}$, which is a measure of complexity of $\mathcal{P}$
\footnote{Specifically, given iid data $Z_1, ..., Z_N$, $Z_i \in \mathcal{Z}$ and a set of functions $\mathcal{F}$ where $f : \mathcal{Z} \to \mathbbm{R}$, the expected Rademacher complexity is defined as
\begin{equation*}
    \mathcal R_N(\mathcal{F}) = \E_{\mathcal{Z}, \sigma}\Big[\sup_f\frac{1}{N}\sum_i^N \sigma_if(\bm{Z}_i)\Big]
\end{equation*}
where $\sigma_1,..., \sigma_N \in \{-1,1\}$ are i.i.d. random variables with $\Prr[\sigma_i = 1]= \Prr[\sigma_i = -1] = 1/2$ \citep[see def. (26.4) in][]{understandingML}.}.
\end{theorem}

The upper bound on the regret depends on three terms. 
Term (a) is the approximation error arising from using the surrogate (approximate) risk instead of the original problem, and it is given by the difference between the optimal surrogate risk and the optimal risk over $\mathcal P$. In other words, this term represents the optimality gap induced by the use of the surrogate loss function.
Terms (b) and (c) control the estimation error of the surrogate risk, $\E_{\mathcal O}[V_{\epsilon,G_0}(\phat{\ipw})-V_{\epsilon,G_0}^\star]$, via the uniform estimation errors $\sup_{p\in\mathcal P}|\Chat(p)-\Cbar{p}|$ and $\sup_{p\in\mathcal P}|\Yhat{\ipw}(p)-\Ybar{p}|$, respectively.
%
In particular, term (c) depends linearly on the expected Rademacher complexity of the policy class multiplied by $n_{\max}^{3/2}$, and is inversely proportional to the lower bound of the propensity score $\eta$ and the margin parameter $\epsilon$.
The convergence properties of this upper bound depend on the convergence to zero of the estimation error and the approximation error. For linear probability policies with bounded covariates and coefficients, $\mathcal R_N(\mathcal{P}_{\Gamma,\mathrm{lin}})=O(N^{-1/2})$. The same rate holds for the class of logistic policies $\mathcal{P}_{\Gamma,\mathrm{logit}}$ under the same conditions.\footnote{By the Contraction Lemma \citep[Lemma 26.9]{understandingML}, the result follows since the logistic function is $1/4$-Lipschitz and the logistic policy class is obtained by composing the linear class with $\sigma(\cdot)$.}
Thus, whenever $\mathcal R_N(\mathcal P)=O(N^{-1/2})$, the estimation error is $O(\epsilon^{-1}N^{-1/2})$. Convergence of the regret bound additionally requires the approximation error to vanish.

On the other hand, the behavior of the approximation error depends only on the margin parameter $\epsilon$. Ideally, if $\epsilon \to 0$, we should expect the approximation error to approach zero as well. Below, we show that, if there exists a direction of growth in the outcome, the approximation error is, in fact, $O(\epsilon)$.

\begin{theorem}[Approximation error]\label{th:approxerr}
For some \(p^\star\in \arg\min_{p\in\mathcal P}R_{G_0}(p)\), \(\Cbar{p}\) is Lipschitz around \(p^\star\), with respect to the norm \(\|\cdot\|\), with Lipschitz constant \(L_p\). Additionally, there exists a direction \(d\) and positive constants $\kappa$ and $t_0$ such that $p^\star+td\in\mathcal P$ and
\(
\Ybar{p^\star+td}\ge \Ybar{p^\star}+\kappa t,
\)
for $t\in[0,t_0]$.
Then, as \(\epsilon \to 0\),
\[
V_{\epsilon,G_0}^\star-R_{G_0}^\star=O(\epsilon).
\]
\end{theorem}

Theorem \ref{th:approxerr} relies on two local properties near the optimal policy $p^\star$: the average treatment probability must be Lipschitz continuous, and there must exist a direction of growth that allows the average outcome to strictly increase. These conditions ensure that a strictly feasible policy can be constructed to satisfy the surrogate's $\epsilon$-buffer without a disproportionate jump in resource costs.
Provided that $R_N(\mathcal{P}) = O(N^{-1/2})$, the regret bound is $O( \epsilon + \epsilon^{-1}N^{-1/2})$. We can then choose the rate of $\epsilon$ optimally by setting $\epsilon \propto N^{-1/4}$, 
%
which leads to a regret bound of order $O(N^{-1/4})$. 

\subsection{Regret Rate Refinement}\label{sec:refinement}
%
The previous result yields a convergence rate of $N^{-1/4}$, which reflects the approximation--estimation tradeoff induced by the surrogate risk, similar to that arising
in surrogate-based policy learning \citep[e.g.,][]{chen2016}. 
%
%
%
%
%
{However, this rate can be improved by exploiting the zero-penalty region of the surrogate loss, provided that the outcome target is attainable with positive slack.
Specifically, define the maximum attainable slack as
\begin{equation}\label{eq:outcomeslack}
s:=\sup_{p\in\mathcal P}\bar Y(p)-T,    
\end{equation}
and, for $\delta\in(0,1)$, let
\[
B_{\ipw}(\delta)
=\frac{M}{\eta}(2n_{\max})^{3/2}\mathcal R_N(\mathcal P)
+\frac{M}{\eta}\sqrt{\frac{2\log(2/\delta)}{N}}.
\]
By Lemma~\ref{lemma:outcome}, $B_{\ipw}(\delta)$ bounds the uniform IPW outcome estimation error with probability at least $1-\delta$.} 
Further, let $c=c(C_0)\in(0,1)$ denote the level at which $h_\epsilon(T+c\,\epsilon;T)=1/C_0$:
\[
c(C_0)=
\begin{cases}
\sqrt{(1-1/C_0)/2}, & 1<C_0\le 2,\\[2pt]
1-(2C_0)^{-1/2}, & C_0\ge 2,
\end{cases}
\]
 
\begin{theorem}[Expected regret refinement]\label{th:raterefinement}
Suppose Assumptions \ref{ass:partial}--\ref{ass:regularity} hold, and that $s>0$. 
For any $\delta\in(0,1)$ and $\epsilon>0$ satisfying
\[
B_{\ipw}(\delta)\le c\epsilon,
\qquad
\epsilon+B_{\ipw}(\delta)<s,
\]
the IPW policy estimator satisfies
\[
\begin{aligned}
\operatorname{Reg}(\hat p_{\ipw})
\le{}&
\underbrace{\bar C^\star\bigl(T+\epsilon+B_{\ipw}(\delta)\bigr)
-\bar C^\star(T)}_{(d)} + \underbrace{4n_{\max}\mathcal R_N(\mathcal P)}_{(e)}
+\underbrace{(1+C_0)\delta}_{(f)}.
\end{aligned}
\]

\end{theorem}
%
{The regret bound of Theorem~\ref{th:raterefinement} has a similar structure to Theorem~\ref{th:regret}, and depends on three terms. Term (d) in the regret bound represents the additional treatment \qq cost" induced by using the surrogate risk with the calibrated buffer $\epsilon \ge  B_{\ipw}(\delta)/c$, and can be interpreted as an approximation error, similarly to term (a) in Theorem~\ref{th:regret}. Term (e) controls estimation error in the average treatment probability, and term (f) bounds the constraint violation probability. The bound thus reflects a related trade-off to that of Theorem~\ref{th:regret} between the cost of ensuring feasibility and the confidence level of that guarantee: decreasing $\delta$ reduces term (f), but increases $B_{\ipw}(\delta)$, requiring a larger $\epsilon$ and potentially increasing term (d).}

{
Theorem~\ref{th:raterefinement} obtains the regret bound by exploiting the zero-penalty region of the surrogate loss, thereby avoiding the $1/\epsilon$ factor arising from uniform control of the surrogate risk. The condition $\epsilon+\Bdev{\ipw}(\delta)<s$ ensures that the buffered target is attainable. On the event that the uniform IPW outcome estimation error is bounded by $\Bdev{\ipw}(\delta)$, which holds with probability at least $1-\delta$, policies attaining this buffered target incur zero empirical penalty. Lemma~\ref{lemma:margin} then implies that $\widehat{\bar Y}^{\ipw}(\phat{\ipw})\ge T+c\epsilon$, so $\epsilon\ge\Bdev{\ipw}(\delta)/c$ guarantees population feasibility. The regret bound follows by comparing treatment costs with those of policies attaining the buffered target $T+\epsilon+\Bdev{\ipw}(\delta)$ (Appendix~\ref{appendix:th3}).}
{
Such regret bound yields a sharper convergence rate. In particular, let $\delta_N=N^{-1}$ and $\epsilon_N=\Bdev{\ipw}(\delta_N)/c$. If $\mathcal R_N(\mathcal P)=O(N^{-1/2})$, then $\epsilon_N+\Bdev{\ipw}(\delta_N)=O(\sqrt{\log N/N})$, so the conditions of Theorem~\ref{th:raterefinement} hold for all $N\ge N_0$, for some $N_0$. Under the additional conditions of Theorem~\ref{th:approxerr}, the approximation error in the average treatment probability is also $O(\sqrt{\log N/N})$, yielding
\[
\operatorname{Reg}(\phat{\ipw})
=O\left(\sqrt{\frac{\log N}{N}}\right).
\]
This suggests choosing the margin \(\epsilon_N\) large enough to dominate the uniform bound \(c^{-1}\Bdev{\ipw}(\delta_N)\). 
In practice, however, the constants entering \(\Bdev{\ipw}(\delta_N)\) may be difficult to obtain or may result in overly conservative margins, producing policies with high treatment probabilities.}
%

\paragraph{Cross-validated margin.}\label{par:cv}
As a practical alternative, we propose a cross-validation procedure to calibrate the finite-sample constant in the margin. 
Let $\epsilon_N(c)=c\sqrt{\log N/ N}$, where
$c\in\mathcal C=\{c_1,\ldots,c_L\}$ is selected from a prespecified grid. We use a cluster-level $K$-fold cross-validation to select $c$, as summarized in Algorithm~\ref{alg:cv-margin}.
\begin{algorithm}[!htbp]
\caption{Cross-validated selection of the IPW margin}
\label{alg:cv-margin}
\begin{algorithmic}[1]
\Require Data from $N$ clusters, target $T$, grid $\mathcal C$, and $K$ folds.
\State Partition the clusters into folds $\mathcal I_1,\ldots,\mathcal I_K$.
\For{$c\in\mathcal C$ and $k=1,\ldots,K$}
    \State Fit on $\mathcal I_{-k}$, containing $N_{-k}$ clusters,
    \[
    \widehat p_c^{(-k)}
    \in\arg\min_{p\in\mathcal P}
    \left\{
    \widehat C_{-k}(p)
    +C_0h_{c/\sqrt{N_{-k}}}
       \bigl(\widehat Y_{-k}^{\ipw}(p);T\bigr)
    \right\}.
    \]
    \State On fold $\mathcal I_k$, compute
    \[
    Y_{k,c}=\widehat Y_k^{\ipw}
       (\widehat p_c^{(-k)}),
    \qquad
    C_{k,c}=\widehat C_k(\widehat p_c^{(-k)}).
    \]
\EndFor
\State For each $c$, calculate
\[
F(c)=\frac1K\sum_{k=1}^K\bm{1}\{Y_{k,c}\ge T\},\quad
S(c)=\frac1K\sum_{k=1}^K(T-Y_{k,c})_+,\quad
R(c)=\frac1K\sum_{k=1}^K C_{k,c}.
\]
\State Select
\[
\widehat c_N
=
\operatorname*{lex\,arg\,min}_{c\in\mathcal C}
\bigl(-F(c),S(c),R(c),c\bigr).
\]
\State \Return $\widehat c_N$.
\end{algorithmic}
\end{algorithm}

Here, $\operatorname{lex\,arg\,min}$
denotes lexicographic minimization, and
$(x)_+=\max\{x,0\}$. For each candidate $c$, the policy is fitted on $K-1$ folds and evaluated on the held-out fold. Candidates are ranked lexicographically, prioritizing the proportion
of folds satisfying the outcome target, followed by the average magnitude of constraint violations and the average treatment probability. Remaining ties are broken in favor of the smaller $c$. The selected constant is then used to refit the policy on the full sample\footnote{Cross-validation and holdout evaluation are standard tools in policy learning for choosing tuning parameters based on estimated out-of-sample performance \citep{zhao2012owl, MbakopTabord2021}. Alternatively, $B_{\ipw}(\delta)$ could be calibrated by a cluster-level multiplier bootstrap approximation to quantiles of the uniform outcome estimation error over the policy class \citep[e.g.][]{kennedy2019}, albeit at a high computational cost. We leave this possibility for future research.}. 

In the simulation study of Section~\ref{sec:simulation}, our cross-validated margin shrinks with the sample size $N$ and the resulting policies attain high feasibility rates with regret decreasing in $N$. In the empirical application of Section~\ref{sec:empirical}, the estimated policies using this procedure are feasible in most held-out folds or fall short of the target only marginally.


\subsection{Regret Bound for the H\'ajek Estimator}\label{sec:hajek-regret}

We next establish uniform finite-sample deviation bounds for the self-normalized H\'ajek estimator over potentially infinite policy classes and use them to show that H\'ajek-based policy learning preserves the regret rates established for IPW.
We restrict attention to policies satisfying $0<p(\bm x,\bm v)<1$, which ensures that the H\'ajek denominator $D_N(p)$ is positive for every policy and sample.\footnote{Restriction to this policy classes is imposed to avoid defining the H\'ajek estimator on zero-denominator samples, and is made for analytical convenience. Policy classes containing 0 and 1 can be accommodated by setting $\Yhat{\hj}(p)=0$ when $D_N(p)=0$, since the centered-moment identity underlying the deviation bounds remain valid under this convention (see Lemma \ref{lem:hajek_uniform}).}
Analogously to $\Bdev{\ipw}(\delta)$ in Section~\ref{sec:refinement}, define, for $\delta \in (0,1)$, the uniform deviation bound
\begin{equation}\label{eq:BH}
\Bdev{\hj}(\delta)=6\sqrt2\frac{M}{\eta}n_{\max}^{3/2}\mathcal R_N(\mathcal P)+\frac{2M}{\eta\sqrt N}+\frac{M}{\eta}\sqrt{\frac{2\log(2/\delta)}{N}}.
\end{equation}
Lemma~\ref{lem:hajek_uniform} then gives $\sup_{p\in\mathcal P}|\Yhat{\hj}(p)-\Ybar{p}|\le\Bdev{\hj}(\delta)$ with probability at least $1-\delta$.

\begin{theorem}[Regret under H\'ajek normalization]\label{th:Hajek1}
Suppose Assumptions~\ref{ass:partial}--\ref{ass:regularity} hold, $0<p(\bm x,\bm v)<1$ for every $p\in\mathcal P$ and $(\bm x,\bm v)\in\mathcal X\times\mathcal V$.

\begin{enumerate}

\item For any $\epsilon>0$,
\begin{align}
\Reg(\phat{\hj})
\leq\;&
V_{\epsilon,G_0}^\star-R_{G_0}^\star
+
8n_{\max}R_N(\mathcal P)
\nonumber
+
24\sqrt{2}\,\frac{C_0M}{\eta\epsilon}n_{\max}^{3/2}\mathcal R_N(\mathcal P)+\frac{4C_0M}{\eta\epsilon\sqrt N}\left(2+\sqrt{2\log2}\right).
\label{eq:hajek-regret-smooth-explicit}
\end{align}

\item Suppose additionally that
\(
s >
0.
\)
For any $\delta\in(0,1)$ and $\epsilon>0$ satisfying
\(
c\epsilon
\geq
{\Bdev{\hj}(\delta)}.
\) and 
\(
\epsilon + {\Bdev{\hj}(\delta)}
< s
\)
it holds:
\begin{align}
\Reg(\phat{\hj})
\leq\;&
\bar C^\star
(
T+\epsilon+\Bdev{\hj}(\delta))
-
\bar C^\star(T)
\nonumber +
8n_{\max}\mathcal R_N(\mathcal P)
+
(1+C_0)\delta.
\end{align}


\end{enumerate}
\end{theorem}

\noindent Theorem~\ref{th:Hajek1} shows that H\'ajek normalization preserves the regret rates obtained with the IPW constraint estimator. In particular, if $\mathcal R_N(\mathcal P)=O(N^{-1/2})$ and the approximation error satisfies the conditions of Theorem~\ref{th:approxerr}, part (i) yields regret $O(\epsilon+\epsilon^{-1}N^{-1/2})$, giving $O(N^{-1/4})$ for $\epsilon\asymp N^{-1/4}$. Part (ii) yields the sharper rate $O(\sqrt{\log N/N})$ by taking $\delta_N=N^{-1}$ and $\epsilon_N=B_{\hj}(\delta_N)/c$. 

{The rate equivalence in Theorem~\ref{th:Hajek1} is expected, paralleling the well-known consistency rate equivalence between H\'ajek and IPW estimators for average treatment effects with and without interference \citep{hajek1971,aronowsamii17}. Nevertheless, H\'ajek normalization can reduce variance and improve finite-sample stability, an advantage not captured by the regret bound rates. Consistent with this, in our simulations under uniform margins, H\'ajek-based policies lead to a lower average treatment probability
while achieving {comparable or higher} feasibility and lower regret than IPW-based policies at every sample size. These gains are not universal, however, as the H\'ajek approach does not uniformly outperform IPW in our empirical application. Similar finite-sample advantages of self-normalization in policy learning without interference have been studied, for instance, by \citet{khanugander2023} and \citet{swaminathan2015selfnormalized}.}

\section{Variance-Adaptive Margin}\label{sec:extension}
In the previous sections, the margin parameter $\epsilon_N$ was chosen uniformly over all policies in $\mathcal{P}$, ignoring the heterogeneous variability of their outcome estimates. 
This uniform approach introduces two sources of inefficiency. First, it over-penalizes low-variance policies whose true performance lies close to the threshold \(T\) but can be estimated with high precision. Second, it fails to adequately guard against high-variance policies, which require wider margins to protect against apparent feasibility driven by sampling noise. 
By ignoring these differences in estimation uncertainty, a uniform margin may lead to unnecessarily high treatment probabilities in finite samples.
To address this issue, we consider a policy-dependent margin that adapts to the estimated variability of the outcome under each candidate policy. For the IPW constraint, the relevant quantity is the sample variance of $\psiipw(p)$, defined in~\eqref{eq:ipwscore}. Then for each $p\in\mathcal{P}$, we set
\[
\widehat\epsilon_N(p)
=
\lambda_N
\sqrt{\frac{\widehat{\Varop}\{\psiipw(p)\}}{N}}
+r_N,
\]
where $\lambda_N$ controls the degree of conservativeness and $r_N\geq 0$ is a lower-order additive correction. The variance-dependent component assigns smaller margins to policies with more precise outcome estimates and larger margins to those with greater uncertainty, potentially reducing resource use while maintaining protection against constraint violations.

This construction is motivated by empirical Bernstein-type generalization bounds \citep{maurerpontil,swaminathan2015crm}, which replace worst-case variance bounds with policy-specific empirical variances. The next theorem establishes regret guarantees for theoretically calibrated $\lambda_N$ and $r_N$. 
In practice, we use the leading variance-dependent term, setting $r_N=0$ and selecting $\lambda_N$ by cross-validation.
\begin{theorem}[Variance-adaptive margin]\label{th:regretboundVA}
Suppose Assumptions~\ref{ass:partial}--\ref{ass:regularity}
hold and $s>0$. For any $N\ge16$ and $\delta\in(0,1)$,
let $\Lambda_N:=\Lambda_{N,\ipw}(\delta/2)$,
defined in Theorem~\ref{th:bernsteinbounds}
of Appendix~\ref{app:lemmas}.

Set $\lambda_N=\sqrt{18\Lambda_N}/c$ and
$r_N=32M\Lambda_N/(c\eta N)$, and for each $p\in\mathcal P$
define the empirical margin
\[
\widehat\epsilon_N(p)
=\lambda_N\sqrt{\frac{\widehat\sigma^2_{\ipw}(p)}{N}}+r_N
=\frac{1}{c}\left\{
\sqrt{\frac{18\,\widehat\sigma^2_{\ipw}(p)\,\Lambda_N}{N}}
+\frac{32M\Lambda_N}{\eta N}
\right\}.
\]
Let
\[
\widehat p^{\,\mathrm{VA}}_{\ipw}
\in\arg\min_{p\in\mathcal P}
\Bigl\{
\widehat{\bar C}(p)
+C_0\,h_{\widehat\epsilon_N(p)}
\bigl(\widehat{\bar Y}_{\ipw}(p);T\bigr)
\Bigr\}.
\]
Define the population buffer and the corresponding
set of buffered-feasible policies,
\[
\Delta_N(p)
=d_1\,\sigma_{\ipw}(p)\sqrt{\frac{\Lambda_N}{N}}
+d_2\,\frac{M}{\eta}\frac{\Lambda_N}{N},
\qquad
\mathcal A^{\Delta}_N
=\bigl\{p\in\mathcal P:\bar Y(p)\ge T+\Delta_N(p)\bigr\},
\]
where $d_1\ge(1+c^{-1})\sqrt{18}$ and
$d_2\ge(1+c^{-1})(16\sqrt{18}+32)$ depend only on $C_0$.
If
\(
\sup_{p\in\mathcal P}\Delta_N(p)<s,
\)
then
\[
\mathrm{Reg}\bigl(\widehat p^{\,\mathrm{VA}}_{\ipw}\bigr)
\le
\underbrace{
\inf_{p\in\mathcal A^\Delta_N}\bar C(p)
-\bar C^\star(T)
}_{\text{(d')}}
+
\underbrace{
4n_{\max}\mathcal R_N(\mathcal P)
}_{\text{(e')}}
+
\underbrace{
(1+C_0)\delta
}_{\text{(f')}}.
\]
\end{theorem}

\noindent The quantities in Theorem~\ref{th:regretboundVA} have the following interpretation. $\sigma^2_{\ipw}(p)=\mathrm{Var}_{G_0}\{\psi_i^{\ipw}(p)\}$ and $\widehat\sigma^2_{\ipw}(p)$ denote the population and sample IPW variances. The term $\Lambda_N$ captures the complexity of $\mathcal P$ through its sup-norm covering numbers $N_\infty(r,\mathcal P,m)$, which measure the number of functions needed to approximate the policy class uniformly over $m$ evaluation points (see Appendix~\ref{app:lemmas}). The theoretical calibration ensures that $c\widehat\epsilon_N(p)$ is a uniform empirical Bernstein confidence radius, satisfying $|\widehat{\bar Y}_{\ipw}(p)-\bar Y(p)|\le c\widehat\epsilon_N(p)$ simultaneously for all $p\in\mathcal P$ with probability at least $1-\delta/2$ (Theorem~\ref{th:bernsteinbounds} in Appendix~\ref{app:lemmas}). The population buffer $\Delta_N(p)$ is constructed to dominate $(1+c)\widehat\epsilon_N(p)$ uniformly with high probability, ensuring that policies satisfying $\bar Y(p)\ge T+\Delta_N(p)$ incur zero empirical surrogate penalty. 
{Consequently, term (d') represents the additional treatment cost induced by the population buffer $\Delta_N(p)$ associated with the policy-specific margin $\widehat\epsilon_N(p)$, analogous to the approximation term in Theorem~\ref{th:raterefinement} for a uniform margin. Terms (e') and (f') are the same as terms (e) and (f) of Theorem~\ref{th:raterefinement}
}

{
To obtain the convergence rate, let $\delta_N=N^{-1}$. For linear or logistic policy classes with bounded covariates and a fixed-dimensional bounded parameter space, $\Lambda_{N,\ipw}(\delta_N/2)=O(\log N)$, so $\sup_{p\in\mathcal P}\Delta_N(p)=O(\sqrt{\log N/N})$\footnote{This follows, for instance, by combining \citet{wainwright}, Lemma 5.7, with the covering-number inequality~\eqref{eq:inequalitycovering} in Theorem~\ref{th:bernsteinbounds} of Appendix~\ref{app:lemmas}.}. Hence, under positive slack, $\mathcal{A}_N^{\Delta}$ is nonempty for all sufficiently large $N$. Combined with $\mathcal R_N(\mathcal P)=O(N^{-1/2})$ and the conditions of Theorem~\ref{th:approxerr}, this yields
\[
\operatorname{Reg}(\widehat p^{\,\mathrm{VA}}_{\ipw})
=O\left(\sqrt{\frac{\log N}{N}}\right).
\]
Although the variance-adaptive procedure achieves the same regret rate as its uniform-margin counterpart, its policy-specific margin can reduce conservativeness and improve finite-sample stability by allowing policies with lower estimated outcome variance to satisfy the constraint with smaller margins. In particular, if $\Delta_N(p)\le\epsilon_N+\Bdev{\ipw}(\delta_N)$ for all $p\in\mathcal P$, the set $\Delta_N(p)$ contains the uniform-margin one, yielding a weakly smaller approximation term (d') compared to (d) in Theorem~\ref{th:raterefinement}. While these potential advantages are not captured by the common convergence rate, our simulations show that variance-adaptive IPW achieves {lower and less variable treatment probabilities and lower regret at every $N$, with comparable feasibility}. In the Colombian application, variance-adaptive IPW similarly reduces average treatment probability across outcome targets.}

The variance-adaptive construction can be extended to the H\'ajek estimator. 
For a policy $p$, define
\[
\widehat{\sigma}_{\hj}^2(p)
=
\frac{1}{N-1}\sum_{i=1}^N
\left\{\widehat{\phi}^{\hj}_i(p)\right\}^2 , \quad \widehat{\phi}^{\hj}_i(p)
= \frac{w_i(p)\{\bar Y_i-\widehat{\bar Y}_{\hj}(p)\}}
     {D_N(p)},
\]
We set the policy-specific margin to
\(
\widehat\epsilon_{N,\hj}(p)
=
\lambda_N\frac{\widehat{\sigma}_{\hj}(p)}{\sqrt N},
\)
and define the following H\'ajek policy estimator: 
\[
\hat{p}^{\mathrm{VA}}_{\hj}\in
\arg\min_{p \in \mathcal{P}} \widehat V^{VA}_{\hj}(p)
, \quad \text{with}\quad 
\widehat V^{VA}_{\hj}(p)
=
\widehat{\bar C}(p)
+
C_0
h_{\widehat\epsilon_{N,\hj}(p)}
\left(
\widehat{\bar Y}_{\hj}(p);T
\right).
\]

The relevant population variance differs from that of the unnormalized IPW
estimator because self-normalization incorporates the covariance between the
weighted outcome and the weights. Consequently, the H\'ajek margin can be
smaller for policies for which normalization reduces estimation variance,
although no uniform ordering between the two estimators is implied.
We establish a uniform empirical Bernstein bound for the Hájek estimator in Appendix~\ref{app:lemmas} (Theorem~\ref{prop:hajek-bernstein}), which yields a corresponding variance-adaptive regret guarantee for $\hat p^{\mathrm{VA}}_{\hj}$ using a theoretically calibrated margin (Theorem \ref{th:hajekva}, Appendix~\ref{app:hajek-bernstein}). The resulting regret bound controls excess resource use relative to policies satisfying a variance-dependent population buffer determined by the H\'ajek estimator variance (see Eq.~\eqref{eq:boundhjva} in Theorem~\ref{th:hajekva}, Appendix~\ref{app:hajek-bernstein}).
\section{Resource-Constrained Problem}\label{sec:dualproblem}
A complementary problem maximizes the population average outcome subject to a constraint on the average treatment probability:
\begin{equation}\label{eq:budgetconstraintproblem}
p^\star \in \arg\max_{p\in\mathcal P}\bar Y(p)
\quad\text{subject to}\quad
\bar C(p)\le B,
\end{equation}
where \(B\) is a prespecified resource budget. Appendix~\ref{app:duality} studies this resource-constrained formulation and establishes equivalence to the outcome-target problem at appropriately matched thresholds under a condition of local outcome growth. This condition implies that any policy leaving resources unused admits an arbitrarily small adjustment that strictly improves the average outcome--a property that holds, for instance, when the average outcome is strictly increasing in treatment.

The two formulations differ in the statistical complexity of their constraints. The resource constraint depends only on observed covariates and policy probabilities, so it is simple to estimate and feasibility may be easier to attain. By contrast, the outcome constraint in \eqref{eq:originalproblem} requires interference-aware estimation of counterfactual outcomes, so ignoring interference can distort the estimated feasible set. In our simulations under the same data-generating process in Appendix~\ref{app:dual_simulation}, the resource-constrained estimators achieve uniformly high feasibility and the no-interference estimator performs comparably to interference-aware methods, whereas in the outcome-target problem the no-interference estimator selects infeasible policies. 
One might think that the resource-constrained problem is easier to learn than the outcome-target problem, since its constraint requires no counterfactual estimation and feasibility is therefore easier to guarantee. However, its objective still requires interference-aware outcome estimation, so its regret guarantees still depend on outcome estimation error.
The simpler estimation of the resource constraint and the equivalence of the two problems at appropriately matched thresholds also suggest that a potential alternative for learning an outcome-target policy is to solve the resource-constrained problem over a grid of budget levels and invert the estimated welfare frontier to identify the minimum $B$ attaining $T$. However, determining this budget level by inverting the estimated welfare frontier remains sensitive to outcome estimation uncertainty. Developing such frontier-inversion procedures and comparing their performance with direct outcome-target learning is an interesting direction for future research.
%


\section{Simulation Study}\label{sec:simulation}

We simulate 500 datasets for each $N\in\{200,400,800,2000\}$, with clusters of size $n_i=10$. For each individual, we generate a single binary covariate $X_{ij}\sim\mathrm{Bernoulli}(0.6)$ and independently assign treatment as $A_{ij}\sim\mathrm{Bernoulli}(0.5)$. 
Let $G_{1ij}=(n_i-1)^{-1}\sum_{k\neq j}X_{ik}A_{ik}$ and $G_{0ij}=(n_i-1)^{-1}\sum_{k\neq j}(1-X_{ik})A_{ik}$
denote the proportions of treated peers with $X=1$ and $X=0$, respectively.
Then, the outcome is generated by
\begin{equation}\label{eq:nonlinear}
Y_{ij}
=
\beta_0
+ \kappa_0 A_{ij}(1-X_{ij})
+ \kappa_1 A_{ij}X_{ij}
+ \theta_1 G_{1ij}(1-A_{ij})X_{ij}
+ \theta_0 G_{0ij}(1-A_{ij})(1-X_{ij})
+ u_{ij},
\end{equation}
where $u_{ij}\sim\mathcal{N}(0,1)$. Note that the terms involving $G_{1ij}$ and $G_{0ij}$ capture heterogeneous spillover exposure within clusters.
We set $\beta_0=0.15$, $\kappa_0=2.25$, $\kappa_1=1.00$, $\theta_0=0.30$, $\theta_1=0.70$, and the outcome target $T=0.75$.
We use the logistic class $\mathcal P_{\Gamma,\mathrm{logit}}$ of Section~\ref{sec:optprobl}, with $p_\gamma(X_{ij}) = \sigma(\gamma_0 + \gamma_1 X_{ij})$ and $\Gamma = [-1,1]^2$, and denote by $\mathcal P_{\Gamma_0,\mathrm{logit}}$, with $\Gamma_0 = [-1,1] \times \{0\}$, its constant subclass.

First, we evaluate four interference-aware estimators combining IPW or H\'ajek outcome estimation with uniform (U) or variance-adaptive (VA) margins: I-IPW-U, I-IPW-VA, I-HJ-U, and I-HJ-VA. Margin constants are selected in each replication by 5-fold cluster-level cross-validation (Algorithm~\ref{alg:cv-margin}). 
Second, we compare the proposed interference-aware estimators with two benchmarks. The first is a no-interference estimator, NI-IPW, which uses the same individualized logistic policy class but replaces the cluster-level IPW estimator with the following IPW estimator ignoring interference
\[
\hat {\bar Y}_{\mathrm{NI-IPW}}(p) = \frac{1}{N}\sum_{i=1}^{N}\frac{1}{n_i}
\sum_{j=1}^{n_i}
Y_{ij}
\frac{\Prr_p(A_{ij}\mid \bm X_{ij},\bm V_i;p)}
{\Prr_{G_0}(A_{ij}\mid \bm X_i,\bm V_i)}.
\]
The second benchmark, the constant-policy estimator CI-IPW, retains the interference-aware IPW estimator but restricts optimization to the constant subclass $\mathcal P_{\Gamma_0,\mathrm{logit}}$.
NI-IPW isolates the consequences of ignoring interference, while CI-IPW assesses the gains from personalizing treatment probabilities.
As population benchmarks, we report the population risk $R_{G_0}(p)$, average treatment probability $\bar C(p)$, and average outcome $\bar Y(p)$ evaluated at the following policies: (i) the optimal personalized policy $p^\star \in \arg\min_{p \in \mathcal P_{\Gamma,\mathrm{logit}}} R_{G_0}(p)$; (ii) the optimal constant policy $q^\star \in \arg\min_{p \in \mathcal P_{\Gamma_0,\mathrm{logit}}} R_{G_0}(p)$; and (iii) the constant policy $\tilde q = \bar C(p^\star)$, which treats every unit uniformly with probability equal to $\bar C(p^\star)$.

\begin{table}[!htbp]
\centering
\caption{Monte Carlo performance of the four interference-aware procedures.}
\label{tab:simulation-margin}
\footnotesize
\setlength{\tabcolsep}{3pt}
\renewcommand{\arraystretch}{0.94}
\begin{threeparttable}
\begin{tabular}{@{}llccccc@{}}
\toprule
$N$ & Estimator & Regret & Treat. Probability & Outcome & Feasibility & Margin \\
\midrule
\multirow{4}{*}{200} & I-IPW-U & 0.160\:(0.119) & 0.461\:(0.070) & 0.965\:(0.105) & 0.984\:(0.126) & 0.2412\: \\
 & I-IPW-VA & 0.100\:(0.066) & 0.414\:(0.027) & 0.857\:(0.032) & 0.996\:(0.063) & 0.1214\: \\
 & I-HJ-U & 0.125\:(0.072) & 0.442\:(0.061) & 0.950\:(0.094) & 0.998\:(0.045) & 0.2052\: \\
 & I-HJ-VA & 0.104\:(0.067) & 0.418\:(0.035) & 0.900\:(0.055) & 0.996\:(0.063) & 0.1596\: \\
\midrule
\multirow{4}{*}{400} & I-IPW-U & 0.108\:(0.082) & 0.420\:(0.047) & 0.913\:(0.074) & 0.994\:(0.077) & 0.1706\: \\
 & I-IPW-VA & 0.084\:(0.078) & 0.396\:(0.024) & 0.839\:(0.026) & 0.994\:(0.077) & 0.0994\: \\
 & I-HJ-U & 0.091\:(0.083) & 0.403\:(0.038) & 0.891\:(0.063) & 0.994\:(0.077) & 0.1431\: \\
 & I-HJ-VA & 0.083\:(0.080) & 0.396\:(0.029) & 0.869\:(0.045) & 0.994\:(0.077) & 0.1242\: \\
\midrule
\multirow{4}{*}{800} & I-IPW-U & 0.088\:(0.115) & 0.392\:(0.034) & 0.872\:(0.056) & 0.986\:(0.118) & 0.1276\: \\
 & I-IPW-VA & 0.080\:(0.126) & 0.381\:(0.023) & 0.826\:(0.024) & 0.984\:(0.126) & 0.0860\: \\
 & I-HJ-U & 0.075\:(0.110) & 0.381\:(0.029) & 0.855\:(0.047) & 0.988\:(0.109) & 0.1079\: \\
 & I-HJ-VA & 0.070\:(0.101) & 0.378\:(0.025) & 0.844\:(0.039) & 0.990\:(0.100) & 0.0989\: \\
\midrule
\multirow{4}{*}{2000} & I-IPW-U & 0.058\:(0.091) & 0.368\:(0.022) & 0.834\:(0.036) & 0.992\:(0.089) & 0.0828\: \\
 & I-IPW-VA & 0.055\:(0.090) & 0.365\:(0.018) & 0.815\:(0.021) & 0.992\:(0.089) & 0.0672\: \\
 & I-HJ-U & 0.045\:(0.049) & 0.362\:(0.019) & 0.823\:(0.032) & 0.998\:(0.045) & 0.0720\: \\
 & I-HJ-VA & 0.046\:(0.066) & 0.361\:(0.018) & 0.820\:(0.029) & 0.996\:(0.063) & 0.0696\: \\
\bottomrule
\end{tabular}
\begin{tablenotes}[flushleft]
\scriptsize
\item \emph{Notes:} Monte Carlo means (standard deviations) over 500 replications. Treat. probability is $\bar C(\hat p)$; Outcome is $\bar Y(\hat p)$; Feasibility is $1\{\bar Y(\hat p)\ge T\}$. Regret refers to $R_{G_0}(\hat p)-R_{G_0}^\star$, where $R(p)=\bar C(p)+1.05\:1\{\bar Y(p)<T\}$, $R_{G_0}^\star=0.3185$, and $T=0.75$.
\end{tablenotes}
\end{threeparttable}
\end{table}

Table~\ref{tab:simulation-margin} summarizes the performance of the four interference-aware procedures. 
The optimal personalized policy attains the target with an average treatment probability of $0.318$. 
For all four procedures, the regret and constraint margins generally decrease with $N$, while feasibility rates remain high ($0.984$--$0.998$).
I-IPW-U has the highest and most variable average treatment probabilities. For instance, at $N=200$ it attains the highest mean outcome ($0.965$) but lowest feasibility ($0.984$), with the highest average treatment probability ($0.461$).
Relative to I-IPW-U, I-IPW-VA uses smaller margins and lower treatment probabilities, with comparable feasibility and lower regret at every $N$. Under uniform margins, Hájek normalization similarly reduces the treatment probability and regret at every $N$, with feasibility at least as high.
The gains from variance-adaptive margins are smaller for the H\'ajek estimator and vanish at $N=2000$. Overall, the proposed procedures perform well, with variance-adaptive margins yielding lower average treatment probabilities with comparable feasibility, particularly when combined with IPW estimation.

\begin{table}[!htbp]
\centering
\caption{Monte Carlo performance of the uniform-margin procedures and benchmarks.}
\label{tab:simulation-comparators}
\footnotesize
\setlength{\tabcolsep}{3pt}
\renewcommand{\arraystretch}{0.94}
\begin{threeparttable}
\begin{tabular}{@{}llccccc@{}}
\toprule
$N$ & Estimator & Regret & Treat. Probability & Outcome & Feasibility & Margin \\
\midrule
\multirow{4}{*}{200} & I-IPW & 0.160\:(0.119) & 0.461\:(0.070) & 0.965\:(0.105) & 0.984\:(0.126) & 0.2412\: \\
 & I-HJ & 0.125\:(0.072) & 0.442\:(0.061) & 0.950\:(0.094) & 0.998\:(0.045) & 0.2052\: \\
 & CI-IPW & 0.112\:(0.027) & 0.430\:(0.027) & 0.869\:(0.041) & 1.000\:(0.000) & 0.1154\: \\
 & NI-IPW & 0.170\:(0.331) & 0.356\:(0.038) & 0.813\:(0.063) & 0.874\:(0.332) & 0.1073\: \\
\midrule
\multirow{4}{*}{400} & I-IPW & 0.108\:(0.082) & 0.420\:(0.047) & 0.913\:(0.074) & 0.994\:(0.077) & 0.1706\: \\
 & I-HJ & 0.091\:(0.083) & 0.403\:(0.038) & 0.891\:(0.063) & 0.994\:(0.077) & 0.1431\: \\
 & CI-IPW & 0.093\:(0.048) & 0.410\:(0.019) & 0.837\:(0.030) & 0.998\:(0.045) & 0.0846\: \\
 & NI-IPW & 0.289\:(0.444) & 0.334\:(0.024) & 0.776\:(0.041) & 0.740\:(0.439) & 0.0755\: \\
\midrule
\multirow{4}{*}{800} & I-IPW & 0.088\:(0.115) & 0.392\:(0.034) & 0.872\:(0.056) & 0.986\:(0.118) & 0.1276\: \\
 & I-HJ & 0.075\:(0.110) & 0.381\:(0.029) & 0.855\:(0.047) & 0.988\:(0.109) & 0.1079\: \\
 & CI-IPW & 0.087\:(0.090) & 0.397\:(0.017) & 0.817\:(0.026) & 0.992\:(0.089) & 0.0671\: \\
 & NI-IPW & 0.530\:(0.512) & 0.319\:(0.017) & 0.751\:(0.030) & 0.496\:(0.500) & 0.0544\: \\
\midrule
\multirow{4}{*}{2000} & I-IPW & 0.058\:(0.091) & 0.368\:(0.022) & 0.834\:(0.036) & 0.992\:(0.089) & 0.0828\: \\
 & I-HJ & 0.045\:(0.049) & 0.362\:(0.019) & 0.823\:(0.032) & 0.998\:(0.045) & 0.0720\: \\
 & CI-IPW & 0.075\:(0.102) & 0.383\:(0.011) & 0.795\:(0.018) & 0.990\:(0.100) & 0.0437\: \\
 & NI-IPW & 0.889\:(0.359) & 0.306\:(0.011) & 0.729\:(0.020) & 0.142\:(0.349) & 0.0339\: \\
\midrule
\multicolumn{7}{@{}l@{}}{\emph{Population benchmarks:}} \\
\multicolumn{7}{@{}l@{}}{\begin{tabular}{@{}lr@{\quad}l@{\quad}l@{\quad}l@{}}
Optimal personalized policy & $p^\star$: & $R_{G_0}(p^\star)=0.3185,$ & $\bar C(p^\star)=0.3185,$ & $\bar Y(p^\star)=0.750$ \\
Optimal constant policy & $q^\star$: & $R_{G_0}(q^\star)=0.354,$ & $\bar C(q^\star)=0.354,$ & $\bar Y(q^\star)=0.750$ \\
Constant policy & $\tilde q$: & $R_{G_0}(\tilde q)=1.3685,$ & $\bar C(\tilde q)=0.3185,$ & $\bar Y(\tilde q)=0.693$
\end{tabular}} \\
\bottomrule
\end{tabular}
\begin{tablenotes}[flushleft]
\scriptsize
\item \emph{Notes:} Monte Carlo means (standard deviations) over 500 replications. Treat. probability is $\bar C(\hat p)$; Outcome is $\bar Y(\hat p)$; Feasibility is $1\{\bar Y(\hat p)\ge T\}$. Regret refers to $R_{G_0}(\hat p)-R_{G_0}^\star$, where $R(p)=\bar C(p)+1.05\:1\{\bar Y(p)<T\}$, $R_{G_0}^\star=0.3185$, and $T=0.75$. Population benchmarks are evaluated exactly under the data-generating process: $p^\star$ is the optimal policy in $\mathcal P_{\Gamma,\mathrm{logit}}$; $q^\star$ is the optimal constant policy in $\mathcal P_{\Gamma_0,\mathrm{logit}}$; and $\tilde q=\bar C(p^\star)=0.3185$ is the constant policy matching the average treatment probability attained at $p^\star$.
\end{tablenotes}
\end{threeparttable}
\end{table}

Table~\ref{tab:simulation-comparators} compares the interference-aware estimators with NI-IPW and CI-IPW, using uniform margins throughout for brevity. Ignoring interference leads to systematically infeasible policies: NI-IPW assigns lower treatment probabilities, but its feasibility rate falls from 0.874 at \(N=200\) to 0.142 at \(N=2000\), where its mean population outcome is 0.729, below target. This occurs because the no-interference estimator evaluates spillover exposure at the experimental treatment rate rather than under the candidate policy, systematically overestimating outcomes for policies with lower treatment probabilities in our design.
In contrast, CI-IPW achieves feasibility comparable to I-IPW. In small samples it also attains lower regret, but for \(N\ge 800\) it requires a higher treatment probability: 0.383 at \(N=2000\), compared with 0.368 for I-IPW. The constant-policy estimator also exhibits less variable average treatment probability and smaller margins, reflecting the lower complexity of the constant policy class.
The population benchmarks isolate the gain from assigning treatment probabilities based on covariates through two complementary comparisons. The optimal personalized policy $p^\star$ attains the target with $\bar C(p^\star) = 0.3185$, whereas the constant policy $\tilde q$, which treats every unit with this same probability and thus has $\bar C(\tilde q) = 0.3185$, achieves only $\bar Y(\tilde q) = 0.693 < T$. The optimal constant policy achieves the outcome target $T$, however it yields $\bar C(q^\star) = 0.354$, 3.6 percentage points more than under $p^\star$.



    



\section{Empirical Application}\label{sec:empirical}
We apply our methodology to the 2005 randomized experiment evaluating the \emph{Subsidios Condicionados a la Asistencia Escolar}, a conditional cash transfer program promoting school attendance and post-secondary enrollment among low-income students in Bogotá, Colombia \citep{osorio}. Students from households with one to five school-aged children were individually randomized within strata defined by locality, school type, gender, and grade. Transfers were conditional on maintaining at least 80 percent monthly school attendance.
This setting creates scope for within-household spillovers. Transfers may encourage siblings' attendance through shared academic environments, peer influence, or parental awareness, but may also divert parental resources toward treated children, reducing attendance among untreated siblings \citep{CamiloZuluaga,ferreira}. Additionally, both direct and spillover effects may be heterogeneous. \citet{osorio} document heterogeneity in the direct response by gender and economic status, and, notably, find that adverse spillovers are asymmetric by gender: untreated female siblings of a treated student experience a sharper attendance decline than untreated male siblings. Spillovers may further depend on household size, since the number of eligible children affects competition for household educational resources \citep{Park2022}.

Motivated by this evidence, we estimate an optimal policy that attains a prespecified aggregate attendance target while accounting for heterogeneous direct and spillover effects. Let $Y_{ij}\in[0,1]$ denote student $j$'s monitored attendance rate in household $i$, and $A_{ij}\in\{0,1\}$ indicate assignment to either cash-transfer intervention. We specify a logistic policy using four covariates: gender, to capture heterogeneous direct and spillover effects; SISBEN socioeconomic status, to capture financial constraints and baseline dropout risk; household size, to capture within-household interference; and locality, to account for differences between experimental sites. Restricting to complete cases yields 6,729 students in 5,725 analytic clusters. We compare I-IPW, I-HJ, NI-IPW, and CI-IPW policies to assess the consequences of ignoring interference and the benefits of personalization.

\paragraph{Out-of-sample evaluation}
We conduct five-fold cross-validation at the household level. We randomly split the households into five folds and, for each fold, estimate the policy on the remaining folds and evaluate it on the held-out fold. Table~\ref{tab:oos_uniform} reports, averaged over folds, the average treatment probability of each learned policy (\emph{Avg.\ prob.}), its interference-aware IPW outcome (\emph{IPW outcome}), the share of folds in which it attains the target (\emph{Feasibility}), and its average shortfall below the target (\emph{Welfare loss}). For comparison, \emph{Own outcome} reports the outcome under each procedure's own estimator. 
\begin{table}[!ht]
\centering
\begin{threeparttable}
\caption{Out-of-sample comparison of I-IPW, I-HJ, NI-IPW, and CI-IPW policies.}\label{tab:oos_uniform}
\scriptsize
\setlength{\tabcolsep}{4pt}
\begin{tabular}{llccccc}
\toprule
Target & Policy & Avg.\ prob. & IPW outcome & Feasibility & Welfare loss & Own outcome \\
\midrule
0.77 & I-IPW & 0.309 & 0.777 & 0.60 & 0.005 & 0.777 \\
0.77 & I-HJ & 0.502 & 0.794 & 0.80 & 0.002 & 0.796 \\
0.77 & NI-IPW & 0.401 & 0.791 & 0.80 & 0.006 & 0.790 \\
0.77 & CI-IPW & 0.532 & 0.794 & 1.00 & 0.000 & 0.794 \\
\midrule
0.78 & I-IPW & 0.518 & 0.799 & 0.80 & 0.003 & 0.799 \\
0.78 & I-HJ & 0.745 & 0.800 & 1.00 & 0.000 & 0.802 \\
0.78 & NI-IPW & 0.620 & 0.806 & 1.00 & 0.000 & 0.803 \\
0.78 & CI-IPW & 0.619 & 0.800 & 1.00 & 0.000 & 0.800 \\
\midrule
0.79 & I-IPW & 0.728 & 0.808 & 1.00 & 0.000 & 0.808 \\
0.79 & I-HJ & 0.865 & 0.806 & 0.80 & 0.001 & 0.806 \\
0.79 & NI-IPW & 0.718 & 0.807 & 0.80 & 0.001 & 0.802 \\
0.79 & CI-IPW & 0.687 & 0.804 & 1.00 & 0.000 & 0.804 \\
\midrule
0.80 & I-IPW & 0.732 & 0.805 & 0.60 & 0.004 & 0.805 \\
0.80 & I-HJ & 0.849 & 0.802 & 0.40 & 0.005 & 0.805 \\
0.80 & NI-IPW & 0.721 & 0.802 & 0.60 & 0.007 & 0.799 \\
0.80 & CI-IPW & 0.731 & 0.807 & 0.80 & 0.001 & 0.807 \\
\midrule
0.81 & I-IPW & 0.777 & 0.811 & 0.60 & 0.004 & 0.811 \\
0.81 & I-HJ & 0.127 & 0.776 & 0.20 & 0.036 & 0.781 \\
0.81 & NI-IPW & 0.272 & 0.778 & 0.20 & 0.034 & 0.780 \\
0.81 & CI-IPW & 0.269 & 0.777 & 0.00 & 0.033 & 0.777 \\
\bottomrule
\end{tabular}
\begin{tablenotes}[flushleft]
\scriptsize
\item \textit{Notes:} Averages over $K = 5$ held-out folds $\mathcal I_k$, with $\hat p^{-k}$ estimated on clusters outside $\mathcal I_k$. \emph{Avg.\ prob.} is $K^{-1}\sum_k \widehat{\bar C}^k(\hat p^{-k})$, where $\widehat{\bar C}^k$ is the average treatment probability on $\mathcal I_k$;\\ \emph{IPW outcome} is $K^{-1}\sum_k \widehat{\bar Y}^k_{\ipw}(\hat p^{-k})$, where $\widehat{\bar Y}^k_{\ipw}$ is the interference-aware IPW outcome estimator on $\mathcal I_k$; \emph{Feasibility} is $K^{-1}\sum_k \bm{1}\{\widehat{\bar Y}^k_{\ipw}(\hat p^{-k}) \ge T\}$; \\ 
\emph{Welfare loss} is $K^{-1}\sum_k [T - \widehat{\bar Y}^k_{\ipw}(\hat p^{-k})]_+$; \emph{Own outcome} is $K^{-1}\sum_k \widehat{\bar Y}^k_m(\hat p^{-k})$, where $\widehat{\bar Y}^k_m$ estimates the outcome on $\mathcal I_k$ by the same estimator in the corresponding procedure, $m \in \{\ipw, \hj, \mathrm{NI\text{-}IPW}\}$.
\end{tablenotes}
\end{threeparttable}
\end{table}

At lower targets, accounting for interference and personalizing treatment probabilities yield substantial resource savings. At $T=0.77$, I-IPW reduces treatment by $9.2$ and $22.3$ percentage points relative to NI-IPW and CI-IPW, respectively, with savings of approximately $10$ percentage points at $T=0.78$. At the latter target, I-IPW and CI-IPW achieve nearly identical outcomes ($0.799$ and $0.800$), confirming the gains from personalization. These savings disappear at $T=0.79$--$0.80$, as meeting higher targets requires treating most individuals, leaving little scope for selective allocation. I-HJ generally remains more treatment-intensive.
Comparing method-specific and interference-aware outcome estimates helps explain these differences. NI-IPW slightly underestimates outcomes at $T=0.77$--$0.80$, potentially contributing to its higher treatment intensity. In contrast, I-HJ's estimates are similar or slightly higher, so outcome underestimation cannot explain its greater resource use. 
The savings under I-IPW come with somewhat lower feasibility: at $T = 0.77$ and $0.78$, its feasibility rate is 0.60 and 0.80, compared with 0.80--1.00 and 1.00 for the other procedures. Nevertheless, the resulting welfare losses are 0.005 and 0.003, and remain below 0.007 across all procedures for targets up to $T = 0.80$.
%
%
At $T=0.81$, I-HJ, NI-IPW, and CI-IPW select low-treatment policies because the target generally cannot be attained in training, leading the optimization to favor minimal treatment. These policies are feasible in at most one fold, with welfare losses of $0.033$--$0.036$, whereas I-IPW remains feasible in three folds, with a welfare loss of $0.004$.

\paragraph{In-sample estimated policy allocations.}


Tables~\ref{tab:profiles} and \ref{tab:marginal-probabilities} report estimated treatment probabilities by covariate profile and averaged within covariate categories, respectively, for I-IPW, I-HJ and NI-IPW.
For covariate $X^k_{ij}$ at value $x^k$, the latter is the weighted average $\hat p_k(x^k) = \sum_{i,j} w_{ij}\,\hat p(\boldsymbol X_{ij}, \boldsymbol V_i) / \sum_{i,j} w_{ij}$, with weights $w_{ij} = n_i^{-1}\bm{1}\{X^k_{ij} = x^k\}$, and analogously for cluster-level covariates.
%
For brevity, we consider $T = (0.77, 0.78, 0.79)$.
Table~\ref{tab:profiles} shows substantial differences across covariate profiles. At $T=0.77$, all three estimators assign the highest probabilities to high-SISBEN males in large households in Suba, and the lowest to low-SISBEN females in small households in San Cristóbal. At $T=0.78$, I-IPW and NI-IPW continue to prioritize males in large households in Suba, whereas I-HJ favors low-SISBEN males in small households in Suba. At $T=0.79$, I-IPW and I-HJ prioritize low-SISBEN females in Suba, in large and small households respectively, while NI-IPW favors low-SISBEN students of either gender in large households in Suba.

Table~\ref{tab:marginal-probabilities} shows which groups receive more treatment on average. All estimators favor Suba across targets. I-IPW and NI-IPW consistently favor larger households, whereas I-HJ favors smaller households at $T=0.78$--$0.79$. At $T=0.77$, all three favor males and high-SISBEN students. At $T=0.78$, males remain favored, but I-HJ and NI-IPW shift toward low-SISBEN students. By $T=0.79$, both interference-aware estimators favor females and low-SISBEN students, whereas NI-IPW assigns nearly equal probabilities by gender. This shift may reflect that prioritizing males and high-SISBEN students is sufficient to meet lower targets, but becomes less effective at higher targets, when adverse spillovers on untreated females and lower attendance among low-SISBEN students may limit attainable outcomes. Meeting these higher targets may therefore require treating more females to mitigate negative spillovers and more low-SISBEN students to improve attendance.

\begin{table}[H]
\centering
\scriptsize
\caption{Treatment probabilities under I-IPW, I-HJ, and NI-IPW by profile.}\label{tab:profiles}
\begin{tabular}{llllccccccccc}
\toprule
 &  &  &  & \multicolumn{3}{c}{$T=0.77$} & \multicolumn{3}{c}{$T=0.78$} & \multicolumn{3}{c}{$T=0.79$}
\\
\cmidrule(lr){5-7} \cmidrule(lr){8-10} \cmidrule(lr){11-13}
Location & SISBEN & Gender & HH size & I-IPW & I-HJ & NI-IPW & I-IPW & I-HJ & NI-IPW & I-IPW & I-HJ & NI-IPW
\\
\midrule
San Cristobal & Low & Female & Small & 0.092 & 0.148 & 0.079 & 0.219 & 0.860 & 0.442 & 0.825 & 0.968 & 0.762 \\
San Cristobal & Low & Female & Large & 0.436 & 0.186 & 0.401 & 0.667 & 0.661 & 0.846 & 0.970 & 0.822 & 0.931 \\
San Cristobal & Low & Male & Small & 0.215 & 0.321 & 0.190 & 0.432 & 0.890 & 0.683 & 0.746 & 0.918 & 0.771 \\
San Cristobal & Low & Male & Large & 0.663 & 0.382 & 0.629 & 0.841 & 0.720 & 0.937 & 0.952 & 0.652 & 0.934 \\
San Cristobal & High & Female & Small & 0.128 & 0.321 & 0.190 & 0.241 & 0.692 & 0.247 & 0.635 & 0.918 & 0.541 \\
San Cristobal & High & Female & Large & 0.518 & 0.381 & 0.625 & 0.691 & 0.429 & 0.697 & 0.922 & 0.657 & 0.832 \\
San Cristobal & High & Male & Small & 0.285 & 0.562 & 0.389 & 0.463 & 0.749 & 0.471 & 0.519 & 0.805 & 0.554 \\
San Cristobal & High & Male & Large & 0.742 & 0.627 & 0.819 & 0.859 & 0.490 & 0.862 & 0.883 & 0.419 & 0.842 \\
\midrule
Suba & Low & Female & Small & 0.647 & 0.759 & 0.610 & 0.836 & 0.969 & 0.935 & 0.982 & 0.982 & 0.977 \\
Suba & Low & Female & Large & 0.916 & 0.799 & 0.903 & 0.968 & 0.921 & 0.988 & 0.997 & 0.923 & 0.994 \\
Suba & Low & Male & Small & 0.833 & 0.896 & 0.809 & 0.932 & 0.976 & 0.975 & 0.972 & 0.954 & 0.978 \\
Suba & Low & Male & Large & 0.969 & 0.916 & 0.964 & 0.989 & 0.934 & 0.996 & 0.995 & 0.795 & 0.994 \\
Suba & High & Female & Small & 0.727 & 0.896 & 0.809 & 0.852 & 0.920 & 0.856 & 0.953 & 0.954 & 0.939 \\
Suba & High & Female & Large & 0.944 & 0.916 & 0.964 & 0.973 & 0.800 & 0.974 & 0.992 & 0.796 & 0.984 \\
Suba & High & Male & Small & 0.879 & 0.959 & 0.920 & 0.940 & 0.938 & 0.942 & 0.926 & 0.883 & 0.942 \\
Suba & High & Male & Large & 0.977 & 0.967 & 0.986 & 0.989 & 0.851 & 0.990 & 0.987 & 0.619 & 0.984 \\
\bottomrule
\end{tabular}
\end{table}

\begin{table}[H]
\centering
\scriptsize
\caption{Marginal implied treatment probabilities for I-IPW, I-HJ, and NI-IPW policies.}\label{tab:marginal-probabilities}
\begin{tabular}{ll*{3}{rrr}}
\toprule
& &
\multicolumn{3}{c}{$T=0.77$} &
\multicolumn{3}{c}{$T=0.78$} &
\multicolumn{3}{c}{$T=0.79$} \\
\cmidrule(lr){3-5}
\cmidrule(lr){6-8}
\cmidrule(lr){9-11}
Dimension & Group
& I-IPW & I-HJ & NI-IPW
& I-IPW & I-HJ & NI-IPW
& I-IPW & I-HJ & NI-IPW \\
\midrule
Gender & Female & 0.297 & 0.336 & 0.324 & 0.441 & 0.744 & 0.542 & 0.824 & 0.894 & 0.761 \\
Gender & Male & 0.452 & 0.518 & 0.481 & 0.620 & 0.779 & 0.716 & 0.753 & 0.777 & 0.764 \\
SISBEN & Low & 0.326 & 0.306 & 0.302 & 0.503 & 0.831 & 0.695 & 0.855 & 0.887 & 0.835 \\
SISBEN & High & 0.421 & 0.544 & 0.501 & 0.555 & 0.691 & 0.561 & 0.723 & 0.786 & 0.690 \\
Household size & Small & 0.278 & 0.427 & 0.308 & 0.430 & 0.825 & 0.539 & 0.730 & 0.910 & 0.710 \\
Household size & Large & 0.611 & 0.421 & 0.634 & 0.778 & 0.601 & 0.850 & 0.937 & 0.652 & 0.894 \\
Location & San Crist\'obal & 0.305 & 0.351 & 0.334 & 0.470 & 0.733 & 0.580 & 0.762 & 0.823 & 0.731 \\
Location & Suba & 0.796 & 0.889 & 0.823 & 0.899 & 0.934 & 0.924 & 0.959 & 0.918 & 0.958 \\
\bottomrule
\end{tabular}
\end{table}

\paragraph{Variance-adaptive margin}
Table~\ref{tab:colombia_final_margin_comparison_oos} compares the uniform and variance-adaptive margins for I-IPW and I-HJ. Here column \emph{Margin} is the average of the cross-validation selected margin across held-out folds. For I-IPW, the adaptive margin is smaller at $T=0.77$--$0.80$ and slightly larger at $T=0.81$; for I-HJ, it is smaller at every target. This pattern reflects the policy-specific variance scaling of the adaptive margin. 
%
For I-IPW, the adaptive margin selects a less resource-intensive policy at every target. Averaging the treatment share by threshold, it reduces the mean treatment share from $0.613$ to $0.514$. 
This reduction is accompanied by somewhat lower out-of-sample feasibility and a slightly higher mean welfare loss. The same trade-off is more pronounced for I-HJ. 
Overall, the variance-adaptive procedure successfully reduces the average treatment probability at the cost of somewhat lower feasibility.
\begin{table}[!htbp]
\centering
\caption{Out-of-sample comparison of uniform and variance-adaptive I-IPW and I-HJ.}
\label{tab:colombia_final_margin_comparison_oos}
\scriptsize
\resizebox{0.85\textwidth}{!}{%
\begin{tabular}{llcccccc}
\toprule
Target & Policy & Avg. prob. & IPW outcome & Feasibility & Welfare loss & Own outcome & Margin
\\
\midrule
0.77 & I-IPW-U & 0.309 & 0.777 & 0.60 & 0.005 & 0.777 & 0.027 \\
0.77 & I-IPW-VA & 0.257 & 0.776 & 0.80 & 0.005 & 0.776 & 0.022 \\
0.77 & I-HJ-U & 0.502 & 0.794 & 0.80 & 0.002 & 0.796 & 0.030 \\
0.77 & I-HJ-VA & 0.127 & 0.776 & 0.60 & 0.009 & 0.781 & 0.007 \\
\midrule
0.78 & I-IPW-U & 0.518 & 0.799 & 0.80 & 0.003 & 0.799 & 0.030 \\
0.78 & I-IPW-VA & 0.365 & 0.782 & 0.60 & 0.006 & 0.782 & 0.020 \\
0.78 & I-HJ-U & 0.745 & 0.800 & 1.00 & 0.000 & 0.802 & 0.026 \\
0.78 & I-HJ-VA & 0.245 & 0.774 & 0.60 & 0.012 & 0.784 & 0.008 \\
\midrule
0.79 & I-IPW-U & 0.728 & 0.808 & 1.00 & 0.000 & 0.808 & 0.028 \\
0.79 & I-IPW-VA & 0.497 & 0.793 & 0.60 & 0.004 & 0.793 & 0.017 \\
0.79 & I-HJ-U & 0.865 & 0.806 & 0.80 & 0.001 & 0.806 & 0.021 \\
0.79 & I-HJ-VA & 0.516 & 0.789 & 0.60 & 0.008 & 0.796 & 0.010 \\
\midrule
0.80 & I-IPW-U & 0.732 & 0.805 & 0.60 & 0.004 & 0.805 & 0.024 \\
0.80 & I-IPW-VA & 0.689 & 0.805 & 0.60 & 0.004 & 0.805 & 0.016 \\
0.80 & I-HJ-U & 0.849 & 0.802 & 0.40 & 0.005 & 0.805 & 0.015 \\
0.80 & I-HJ-VA & 0.714 & 0.802 & 0.40 & 0.006 & 0.799 & 0.006 \\
\midrule
0.81 & I-IPW-U & 0.777 & 0.811 & 0.60 & 0.004 & 0.811 & 0.010 \\
0.81 & I-IPW-VA & 0.763 & 0.810 & 0.60 & 0.005 & 0.810 & 0.011 \\
0.81 & I-HJ-U & 0.127 & 0.776 & 0.20 & 0.036 & 0.781 & 0.017 \\
0.81 & I-HJ-VA & 0.127 & 0.776 & 0.20 & 0.036 & 0.781 & 0.007 \\
\bottomrule
\end{tabular}
}
\end{table}

Appendix~\ref{appendix:additional} reports the treatment probabilities by covariate category under variance-adaptive margins: as with uniform margins, I-IPW consistently assigns higher probabilities to larger households, whereas I-HJ favors smaller households at $T = 0.78$--$0.80$.



\section{Conclusion}\label{sec:conclusion}
This paper develops a framework for learning resource-efficient individualized stochastic treatment policies under heterogeneous partial interference. By combining interference-aware estimation with a smooth approximation to the outcome constraint, we obtain policies that balance resource use and population feasibility, together with non-asymptotic regret guarantees for both IPW and self-normalized estimators. Our variance-adaptive margin further accounts for policy-specific estimation uncertainty and can substantially reduce resource use. Our simulation study and the Colombian cash-transfer application illustrate the practical importance of accounting for interference and treatment heterogeneity when designing policies to achieve prespecified population-level targets.

Our framework suggests three potential extensions. First, policy estimation could incorporate observed network information. This applies both under partial interference, with networks observed within clusters, and in a single interconnected network, where estimation and regret bounds must also account for network dependence \citep{leung2020treatment,viviano2024}. 
Second, the framework could be extended to observational studies with unknown propensity scores, using flexible doubly robust estimators, requiring guarantees that account for nuisance estimation error \citep{LeeZengHudgens,park24}. Finally, building on \citet{park24}, the framework could be extended to individualized policies subject to cluster-level outcome targets rather than a population-level target. This would require adapting the constraint surrogate to the cluster level, allowing personalized treatment allocation while enforcing cluster-specific outcome requirements.

\bibliography{bibliography}

\newpage
\appendix
\counterwithin{theorem}{section}
\counterwithin{lemma}{section}
\counterwithin{assumption}{section}
\counterwithin{equation}{section}

\begin{bibunit}

\section{Resource-Constrained Optimal Policies}\label{appendix:dualproblem}

\subsection{The dual resource-constrained problem}\label{app:duality}

The main outcome-target problem formulation minimizes the average treatment probability subject to a population outcome target. The complementary resource-constrained problem maximizes the average outcome over a policy class under a constraint on the average treatment probability 
\begin{equation}\label{eq:dual_budget}
  \max_{p\in\mathcal P}\ \bar Y(p)
  \quad\text{subject to}\quad \bar C(p)\le B.
\end{equation}
with $B \in [0,1]$. Consistent with the main text, define the population value functions
\begin{equation}\label{eq:dual_frontiers}
 \bar C^\star(T)=\inf_{p\in\mathcal P:\,\bar Y(p)\ge T}\bar C(p),
 \qquad
 \bar Y^\star(B)=\sup_{p\in\mathcal P:\,\bar C(p)\le B}\bar Y(p).
\end{equation}

We next characterize the relationship between the outcome-target problem in Eq.~\ref{eq:minproblem} and the resource-constrained problem in Eq.~\ref{eq:dual_budget}. Given a budget $B$, consider the outcome-target problem with target $T^\star = \bar Y^\star(B)$, the maximal outcome attainable under $B$. Every budget-constrained maximizer attains $T^\star$ with average treatment probability at most $B$, so $\bar C^\star(T^\star) \le B$. Equality, and hence identical optimizer sets, holds if and only if every budget-constrained maximizer exhausts the budget: a maximizer $p$ with $\bar C(p) < B$ attains $T^\star$ with a strictly lower average treatment probability, so the two problems have different solutions. The following condition rules this out.


\begin{assumption}[Local outcome growth]\label{ass:localgrowth}
Fix a budget $B$ for which the outcome-maximization problem has a solution.
For every $p\in\mathcal P$ with $\bar C(p)<B$ and every $\varepsilon>0$, there exists $p'\in\mathcal P$ such that
\begin{equation}\label{eq:dual_nonsat}
 \lVert p'-p\rVert_\infty<\varepsilon,
 \qquad \bar Y(p')>\bar Y(p).
\end{equation}
\end{assumption}
\noindent Assumption \ref{ass:localgrowth} is a local outcome growth condition on the attainable stochastic policies; it does not require welfare to be globally increasing in every treatment
probability.

\begin{theorem}[Equivalence at matched thresholds]\label{thm:dual_equiv}
Suppose Assumption~\ref{ass:localgrowth} holds and the budget-constrained problem
\eqref{eq:dual_budget} admits an optimizer. Set $T^\star=\bar Y^\star(B)$.
Then $\bar C^\star(T^\star)=B$, and
\[
 \operatorname*{arg\,max}_{p\in\mathcal P:\,\bar C(p)\le B}\bar Y(p)
 =
 \operatorname*{arg\,min}_{p\in\mathcal P:\,\bar Y(p)\ge T^\star}\bar C(p).
\]
\end{theorem}

Assumption~\ref{ass:localgrowth} is sufficient, not necessary. For a fixed
budget $B$ with attained optima, $\bar C^\star\{\bar Y^\star(B)\}=B$ holds if and only if every
maximizer under budget $B$ exhausts that budget. One sufficient primitive
condition for local outcome growth is that every nontrivial componentwise
increase in a cluster treatment vector strictly improves its cluster-average
potential outcome, almost surely, \emph{together with} a policy class that
allows arbitrarily small, nontrivial componentwise upward perturbations
on a set of clusters of positive probability whenever $\bar C(p)<B$. This condition concerns aggregate effects, so individual
negative spillovers are permitted when outweighed by positive effects on
other cluster members. By contrast, sufficiently adverse spillovers may
create welfare-satiation regions.

For illustration, consider a cluster containing two siblings, with
\[
 Y_{i1}(a_1,a_2)=0.60+0.20a_1-0.25a_2,\qquad
 Y_{i2}(a_1,a_2)=0.60+0.15a_2.
\]
For independent treatment probabilities $p=(p_1,p_2)$, the expected
cluster-average outcome and intensity are
\[
 \bar Y(p)=0.60+0.10p_1-0.05p_2,\qquad
 \bar C(p)=\frac{p_1+p_2}{2}.
\]
Restrict $p_1,p_2\in[0.2,0.8]$ and take $B=0.75$.
The budget-constrained maximizer is $p^\star=(0.8,0.2)$, with
$\bar Y^\star(0.75)=0.67$ but $\bar C(p^\star)=0.50$. Consequently,
$\bar C^\star\{\bar Y^\star(0.75)\}=0.50<0.75$. Thus the value-function inversion at
$T=\bar Y^\star(B)$ and equality of optimizer sets can fail when a budget-constrained
optimum leaves resources unused, even though the general relation
$\bar C^\star(T)=\inf\{b:\bar Y^\star(b)\ge T\}$ continues to hold.

\subsubsection{Estimation and regret for outcome maximization}\label{app:dual_estimation}

Let $\widehat{\bar Y}^{\ipw}(p)$ and $\widehat{\bar C}(p)$ be
as in the main text. Define the reflected smooth hinge
\[
 h_\epsilon^+(c;B)=h_\epsilon(-c;-B).
\]
It is zero for $c\le B-\epsilon$, equals one for $c\ge B$, and is
$2/\epsilon$-Lipschitz. Fix a penalty level $C_y>2M$, where
$|Y_{ij}(a_i)|\le M$, and define
\begin{align*}
 V^W_{\epsilon}(p)
   &=\bar Y(p)-C_y h_\epsilon^+\{\bar C(p);B\},\\
 \widehat V^W_{\epsilon}(p)
   &=\widehat{\bar Y}^{\ipw}(p)
        -C_y h_\epsilon^+\{\widehat{\bar C}(p);B\},\\
 \widehat p^W_\epsilon
   &\in\operatorname*{arg\,max}_{p\in\mathcal P}
          \widehat V^W_\epsilon(p).
\end{align*}
The population penalized criterion and expected regret are
\[
 R^W(p)=\bar Y(p)-C_y\bm{1}\{\bar C(p)>B\},\qquad
 \mathrm{Reg}^W(\widehat p)=\bar Y^\star(B)-\mathbb E\{R^W(\widehat p)\}.
\]
Since $C_y>2M$ and there is a budget-feasible policy,
$\sup_{p\in\mathcal P}R^W(p)=\bar Y^\star(B)$.

\paragraph{Budget-side approximation error.}
An analogue of Theorem~2 controls the surrogate approximation error through
a local direction of decreasing resource use. Let $p_B^\star$ attain
$\bar Y^\star(B)$. If $\bar C(p_B^\star)<B$, then
$\bar Y^\star(B-\epsilon)=\bar Y^\star(B)$ for all sufficiently small
$\epsilon>0$. Otherwise, suppose $\bar C(p_B^\star)=B$ and that there
are constants $\kappa,L,t_0>0$ and a path $\{p_t:0\le t\le t_0\}
\subseteq\mathcal P$ satisfying
\begin{equation}\label{eq:dual_downward_direction}
 p_0=p_B^\star,\qquad
 \bar C(p_t)\le B-\kappa t,\qquad
 \lVert p_t-p_B^\star\rVert_\infty\le Lt
 \quad(0\le t\le t_0).
\end{equation}
Under Assumption~3, the population outcome functional is already
Lipschitz with respect to the policy probabilities:
\begin{equation}\label{eq:dual_welfare_lipschitz}
 |\bar Y(p)-\bar Y(q)|\le 2Mn_{\max}\lVert p-q\rVert_\infty.
\end{equation}
Indeed, couple each cluster's independent Bernoulli treatment vectors
under $p$ and $q$ using the same independent uniforms. The probability
that any treatment differs is at most
$\sum_{j=1}^{n_i}|p(X_{ij},V_i)-q(X_{ij},V_i)|$, while a cluster-average
potential outcome changes by at most $2M$ if the assignment vector
changes. Taking expectations gives \eqref{eq:dual_welfare_lipschitz}.
For $0<\epsilon\le\kappa t_0$, take $t=\epsilon/\kappa$.
Then $\bar C(p_t)\le B-\epsilon$ and
\begin{equation}\label{eq:dual_frontier_lipschitz}
 0\le\bar Y^\star(B)-\bar Y^\star(B-\epsilon)
 \le\bar Y(p_B^\star)-\bar Y(p_t)
 \le\frac{2Mn_{\max}L}{\kappa}\epsilon.
\end{equation}
Hence the approximation error is $O(\epsilon)$ without assuming that
$\mathcal P$ contains the zero-treatment policy or is closed under
probability scaling.

Condition \eqref{eq:dual_downward_direction} is readily verified for a
logistic policy family when the parameter space permits a small downward
shift of the intercept. Writing
$p_\gamma(x,v)=\sigma\{\gamma_0+g_{\widetilde\gamma}(x,v)\}$,
set $p_t=p_{(\gamma_0^\star-t,\widetilde\gamma^\star)}$.
Since $\sigma$ is $1/4$-Lipschitz,
$\lVert p_t-p_B^\star\rVert_\infty\le t/4$; furthermore,
\[
 \left.\frac{\mathrm d}{\mathrm dt}\bar C(p_t)\right|_{t=0}
 =-\mathbb E_{G_0}\left\{\frac1{n_i}\sum_{j=1}^{n_i}
 \sigma'\bigl(\gamma_0^\star+g_{\widetilde\gamma^\star}(X_{ij},V_i)\bigr)\right\}<0.
\]
For finite predictors, continuity then provides a $\kappa>0$ on a
sufficiently short interval. Mere continuity of the policy function
is not sufficient: the admissible class must allow the downward
perturbation. Unlike the outcome-growth condition of Theorem~2,
this property depends on the policy class rather than on the sign of
direct or spillover treatment effects.

\begin{theorem}[Regret for resource-constrained policies]
\label{thm:dual_regret}
Suppose Assumptions 1--3 hold, $C_y>2M$, and the relevant
optimization problems admit solutions.
Let $\mathfrak R_N=\mathfrak R_N(\mathcal P)$.

\emph{(i) General surrogate bound.}
For $\epsilon>0$ such that $\bar Y^\star(B-\epsilon)$ is defined,
\begin{align}\label{eq:dual_general_regret}
 \mathrm{Reg}^W(\widehat p^W_\epsilon)
 &\le \bar Y^\star(B)-\bar Y^\star(B-\epsilon)
 +4\sqrt2\,\frac{M}{\eta}n_{\max}^{3/2}\mathfrak R_N
 +\frac{2M}{\eta}\sqrt{\frac{2\log 2}{N}}
 +\frac{16C_y n_{\max}}{\epsilon}\mathfrak R_N.
\end{align}
If the local downward-direction condition
\eqref{eq:dual_downward_direction} holds and
$\mathfrak R_N=O(N^{-1/2})$, then choosing
$\epsilon\asymp N^{-1/4}$ gives
$\mathrm{Reg}^W(\widehat p^W_\epsilon)=O(N^{-1/4})$.

\emph{(ii) Refined bound.}
For $\delta\in(0,1)$ define
\begin{align*}
 b_C(\delta)
  &=2n_{\max}\mathfrak R_N
      +\sqrt{\frac{\log(4/\delta)}{2N}},\\
 b_Y(\delta)
  &=2\sqrt2\frac{M}{\eta}n_{\max}^{3/2}\mathfrak R_N
      +\frac{M}{\eta}\sqrt{\frac{2\log(4/\delta)}{N}},\\
 \alpha_N(\delta)
  &=\frac{2M+2b_Y(\delta)}{C_y},
\end{align*}
and let
\[
 k(\alpha)=
 \begin{cases}
    \sqrt{(1-\alpha)/2},& 1/2\le\alpha<1,\\
    1-\sqrt{\alpha/2},& 0<\alpha<1/2.
 \end{cases}
\]
Suppose $\alpha_N(\delta)<1$, that
$\bar Y^\star\{B-\epsilon-b_C(\delta)\}$ is defined, and that
\(
 \epsilon\,k\{\alpha_N(\delta)\}\ge b_C(\delta).
\)
Then
\begin{align}\label{eq:dual_refined_regret}
 \mathrm{Reg}^W(\widehat p^W_\epsilon)
 &\le
 \bar Y^\star(B)-\bar Y^\star\{B-\epsilon-b_C(\delta)\}
 +2b_Y(\delta)+(2M+C_y)\delta.
\end{align}
If the local downward-direction condition
\eqref{eq:dual_downward_direction} holds,
$\mathfrak R_N=O(N^{-1/2})$, and $C_y>2M$ is fixed, then
$\delta_N=N^{-1}$ and a choice
$\epsilon_N\asymp\sqrt{\log N/N}$ sufficiently large to satisfy
the displayed margin condition yield
$\mathrm{Reg}^W(\widehat p^W_{\epsilon_N})
 =O\{\sqrt{\log N/N}\}$.
\end{theorem}

The distinction between the two regret regimes parallels the main-text
analysis. The general bound trades the left-hand frontier approximation
error $\bar Y^\star(B)-\bar Y^\star(B-\epsilon)$ against an estimation term of order
$\epsilon^{-1}\mathfrak R_N$. The refined bound instead uses uniform
constraint concentration to certify the population budget feasibility of
the learned policy. Importantly, it also incurs outcome-estimation error
through the \emph{objective}. Hence its favorable feasibility guarantee
does not imply that learning the welfare-maximizing policy is free of
interference-related statistical uncertainty. A H\'ajek analogue can be obtained
under the strict-positivity conditions used in the main text by replacing
the IPW deviation bound with the corresponding self-normalized bound.

\subsection{Simulation study}\label{app:dual_simulation}

We use the data-generating process and individualized logistic policy
class of Section~7. The expected treatment-intensity budget is
$B=0.3185$, matching the oracle treatment share associated with the
outcome target $T=0.75$ in the main simulation. We examine
interference-aware IPW and H\'ajek estimators with a uniform hinge
margin, together with the no-interference IPW estimator (NI-IPW) and a
constant-policy interference-aware estimator (CI-IPW). The margin
$\epsilon_N=0.1\sqrt{\log(N)/N}$ is held common across methods.
\begin{table}[!htbp]
\centering
\caption{Monte Carlo performance of the budget-constrained uniform-margin procedures and benchmarks.}
\label{tab:simulation-budget}
\footnotesize
\setlength{\tabcolsep}{3pt}
\renewcommand{\arraystretch}{0.94}
\begin{threeparttable}
\begin{tabular}{@{}llccccc@{}}
\toprule
$N$ & Estimator & Regret & Cost & Outcome & Feasible & Margin \\
\midrule
\multirow{4}{*}{200} & I-IPW & 0.059\:(0.051) & 0.295\:(0.027) & 0.691\:(0.051) & 1.000\:(0.000) & 0.0163 \\
 & I-HJ & 0.043\:(0.030) & 0.299\:(0.015) & 0.707\:(0.030) & 1.000\:(0.000) & 0.0163 \\
 & CI-IPW & 0.086\:(0.009) & 0.301\:(0.005) & 0.664\:(0.009) & 1.000\:(0.000) & 0.0163\\
 & NI-IPW & 0.038\:(0.006) & 0.301\:(0.003) & 0.712\:(0.006) & 1.000\:(0.000) & 0.0163 \\
\midrule
\multirow{4}{*}{400} & I-IPW & 0.042\:(0.041) & 0.301\:(0.022) & 0.708\:(0.041) & 1.000\:(0.000) & 0.0122 \\
 & I-HJ & 0.032\:(0.018) & 0.304\:(0.008) & 0.718\:(0.018) & 1.000\:(0.000) & 0.0122 \\
 & CI-IPW & 0.079\:(0.004) & 0.305\:(0.003) & 0.671\:(0.004) & 1.000\:(0.000) & 0.0122 \\
 & NI-IPW & 0.031\:(0.005) & 0.305\:(0.002) & 0.719\:(0.005) & 1.000\:(0.000) & 0.0122\\
\midrule
\multirow{4}{*}{800} & I-IPW & 0.031\:(0.027) & 0.306\:(0.014) & 0.719\:(0.027) & 1.000\:(0.000) & 0.0091\\
 & I-HJ & 0.027\:(0.013) & 0.307\:(0.006) & 0.723\:(0.013) & 1.000\:(0.000) & 0.0091 \\
 & CI-IPW & 0.074\:(0.001) & 0.308\:(0.001) & 0.676\:(0.001) & 1.000\:(0.000) & 0.0091 \\
 & NI-IPW & 0.027\:(0.005) & 0.308\:(0.002) & 0.723\:(0.005) & 1.000\:(0.000) & 0.0091\\
\midrule
\multirow{4}{*}{2000} & I-IPW & 0.023\:(0.007) & 0.311\:(0.002) & 0.727\:(0.007) & 1.000\:(0.000) & 0.0062\\
 & I-HJ & 0.024\:(0.006) & 0.310\:(0.002) & 0.726\:(0.006) & 1.000\:(0.000) & 0.0062 \\
 & CI-IPW & 0.069\:(0.002) & 0.311\:(0.001) & 0.681\:(0.002) & 1.000\:(0.000) & 0.0062 \\
 & NI-IPW & 0.022\:(0.004) & 0.310\:(0.002) & 0.728\:(0.004) & 1.000\:(0.000) & 0.0062\\
\bottomrule
\end{tabular}
\begin{tablenotes}[flushleft]
\scriptsize
\item \emph{Notes:} Monte Carlo means (standard deviations) over 500 replications. Cost is $\bar C(\hat p)$; Outcome is $\bar Y(\hat p)$; Feasible is $1\{\bar C(\hat p)\le B\}$. Regret is $R^{W\star}_{G_0}-R^W_{G_0}(\hat p)$, where $R^W(p)=\bar Y(p)-C_y1\{\bar C(p)>B\}$, $C_y=50$, $R^{W\star}_{G_0}=0.7500$, and $B=0.3185$. The uniform margin is $0.1\sqrt{\log(N)/N}$. The reference is the population oracle at budget $B$, without subtracting the finite-sample margin.
\end{tablenotes}
\end{threeparttable}
\end{table}

All four procedures satisfy the \emph{population} expected-resource
constraint in every one of the 500 Monte Carlo replications for each
sample size considered. This is consistent with the simpler structure
of the expected treatment-intensity constraint, which depends only on
observed covariates and policy probabilities. The feasibility finding
must, however, be interpreted alongside welfare: at $N=2000$ the
average outcomes under interference-aware IPW and H\'ajek are $0.727$
and $0.726$, respectively, below the oracle value $\bar Y^\star(B)=0.750$.
The corresponding average treatment intensities, $0.311$ and $0.310$,
are also below the budget $B=0.3185$. Part of the welfare loss
therefore reflects conservative use of the available expected budget.
As the sample size grows, treatment intensity approaches $B$ and
welfare regret generally declines.

A second difference from the outcome-target formulation concerns
misspecification of interference. In the present budget-constrained
simulation, NI-IPW attains welfare close to the interference-aware
procedures: at $N=2000$, its mean outcome is $0.728$, versus $0.727$
and $0.726$ for interference-aware IPW and H\'ajek. This result suggests
that, in this design and at this budget, ignoring interference does not
substantially change the population welfare attained by the selected
policy. 
The distinction is more pronounced in the outcome-target simulation of
Section~7, where NI-IPW incorrectly classifies policies as meeting the
outcome target and its population feasibility rate falls to {$0.142$} at
$N=2000$. In budget-constrained welfare maximization, ignoring
interference biases the estimated \emph{objective} but does not bias the
expected-intensity constraint; good policy selection remains possible
if the misspecified objective preserves enough information about the
relative welfare of competing policies. In outcome-constrained resource
minimization, the same misspecification directly distorts the
\emph{estimated feasible set}, potentially causing systematic target
violations. 

Finally, CI-IPW achieves an average outcome of $0.681$ at $N=2000$,
compared with $0.727$ for individualized interference-aware IPW,
illustrating the welfare gains from covariate-based personalization
under the same expected-resource budget.

\subsection{Colombian CCT application under a budget constraint}
\label{app:dual_colombia}
Table~8 reports held-out performance of the budget-constrained estimators
for budgets $B\in\{0.35,0.50,0.65,0.80\}$; 
All methods satisfy the held-out expected-resource constraint in every
reported outer fold, with no observed budget excess. Higher budgets are
associated with greater treatment intensity and generally higher
estimated outcomes: for interference-aware IPW, the treatment share rises
from $0.328$ to $0.768$, while its held-out IPW-evaluated outcome rises
from $0.783$ to $0.812$. At a given budget, the estimated outcomes of
the individualized interference-aware and no-interference policies are
fairly similar under the common IPW evaluation column, consistent with
the simulation findings. These are estimated, not known population,
outcomes; the comparison therefore does not show that interference can
be ignored generally. 

\begin{table}[!htbp]
\centering
\footnotesize
\resizebox{0.90\textwidth}{!}{%
\begin{tabular}{llccccc}
\toprule
Budget & Policy & Treat. Prob & Own outcome & I-IPW outcome & Budget feasibility \\
\midrule
0.35 & I-IPW & 0.328 & 0.783 & 0.783 & 1.00 \\
0.35 & I-HJ & 0.328 & 0.793 & 0.785 & 1.00 \\
0.35 & NI-IPW & 0.328 & 0.790 & 0.790 & 1.00 \\
0.35 & CI-IPW & 0.329 & 0.780 & 0.780 & 1.00 \\
\midrule
0.50 & I-IPW & 0.469 & 0.799 & 0.799 & 1.00 \\
0.50 & I-HJ & 0.477 & 0.797 & 0.788 & 1.00 \\
0.50 & NI-IPW & 0.462 & 0.793 & 0.797 & 1.00 \\
0.50 & CI-IPW & 0.479 & 0.790 & 0.790 & 1.00 \\
\midrule
0.65 & I-IPW & 0.587 & 0.802 & 0.802 & 1.00 \\
0.65 & I-HJ & 0.627 & 0.798 & 0.796 & 1.00 \\
0.65 & NI-IPW & 0.580 & 0.796 & 0.800 & 1.00 \\
0.65 & CI-IPW & 0.629 & 0.800 & 0.800 & 1.00 \\
\midrule
0.80 & I-IPW & 0.768 & 0.812 & 0.812 & 1.00 \\
0.80 & I-HJ & 0.777 & 0.803 & 0.807 & 1.00 \\
0.80 & NI-IPW & 0.774 & 0.807 & 0.811 & 1.00 \\
0.80 & CI-IPW & 0.779 & 0.810 & 0.810 & 1.00 \\
\bottomrule
\end{tabular}
}
\caption{Out-of-sample Colombia budget-constraint results under uniform margins. Feasibility is the fraction of outer folds satisfying the budget.}
\label{tab:colombia-budget-oos}
\end{table}

\section{Additional Tables}\label{appendix:additional}
\begin{table}[H]
\centering
\caption{Marginal implied treatment probabilities under I-IPW-VA and I-HJ-VA.}
\label{tab:colombia_probability_marginals_adaptive}
\footnotesize
\resizebox{0.90\textwidth}{!}{%
\begin{tabular}{lllrrrrr}
\toprule
Dimension & Group & Policy  type & T=0.77 & T=0.78 & T=0.79 & T=0.80 & T=0.81 \\
\midrule
Gender & Female & I-IPW-VA & 0.244 & 0.346 & 0.368 & 0.767 & 0.836 \\
Gender & Female & I-HJ-VA & 0.099 & 0.186 & 0.390 & 0.522 & 0.099 \\
Gender & Male & I-IPW-VA & 0.399 & 0.499 & 0.547 & 0.772 & 0.761 \\
Gender & Male & I-HJ-VA & 0.216 & 0.329 & 0.592 & 0.720 & 0.216 \\
SISBEN & Low SISBEN & I-IPW-VA & 0.329 & 0.402 & 0.430 & 0.843 & 0.864 \\
SISBEN & Low SISBEN & I-HJ-VA & 0.104 & 0.164 & 0.367 & 0.641 & 0.104 \\
SISBEN & High SISBEN & I-IPW-VA & 0.312 & 0.440 & 0.483 & 0.695 & 0.734 \\
SISBEN & High SISBEN & I-HJ-VA & 0.209 & 0.349 & 0.611 & 0.598 & 0.209 \\
HH size & Small & I-IPW-VA & 0.196 & 0.304 & 0.352 & 0.711 & 0.744 \\
HH size & Small & I-HJ-VA & 0.095 & 0.332 & 0.508 & 0.653 & 0.095 \\
HH size & Large & I-IPW-VA & 0.632 & 0.716 & 0.718 & 0.915 & 0.937 \\
HH size & Large & I-HJ-VA & 0.311 & 0.068 & 0.443 & 0.536 & 0.311 \\
Location & San Cristobal & I-IPW-VA & 0.368 & 0.402 & 0.402 & 0.746 & 0.775 \\
Location & San Cristobal & I-HJ-VA & 0.163 & 0.207 & 0.430 & 0.566 & 0.163 \\
Location & Suba & I-IPW-VA & 0.027 & 0.542 & 0.796 & 0.910 & 0.947 \\
Location & Suba & I-HJ-VA & 0.117 & 0.564 & 0.856 & 0.953 & 0.117 \\
\bottomrule
\end{tabular}
}
\end{table}


\section{Proofs}\label{appendix:proof}
This section provides the derivations of Theorem \ref{th:regret}-\ref{th:regretboundVA} in the main text, as well as Theorem \ref{th:hajekva} for the regret bound on the Hajek policy estimator with variance-adaptive margin, and the results in Appendix~\ref{appendix:dualproblem}.
We start by introducing some preliminary lemmas, and then move to the main proofs.

\subsection{Preliminary Lemmas and Propositions}\label{app:lemmas}

\begin{lemma}[Uniform deviation bounds for the average treatment probability]\label{lemma:cost}
Suppose Assumption \ref{ass:regularity}.\ref{ass:boundedclus} holds. For any $\delta\in(0,1)$, with probability at least $1-\delta$,
\[
\sup_{p\in\mathcal P}|\Cbar{p}-\Chat(p)|\le 2n_{\max}\mathcal R_N(\mathcal P)+\sqrt{\frac{\log(2/\delta)}{2N}}.
\]
Moreover,
\[
\E_{\mathcal O}\left[\sup_{p\in\mathcal P}|\Cbar{p}-\Chat(p)|\right]\le 4n_{\max}\mathcal R_N(\mathcal P).
\]
\end{lemma}

\begin{proof}
Let $f_p(\bm O_i)=n_i^{-1}\sum_{j=1}^{n_i}p(\bm X_{ij},\bm V_i)$ and $\mathcal F'=\{f_p:p\in\mathcal P\}$. Then $0\le f_p\le1$, $\Chat(p)=N^{-1}\sum_i f_p(\bm O_i)$, and $\Cbar{p}=\E_{G_0}f_p(\bm O_i)$. The one-sided symmetrization inequality \citep[Lemma 26.2]{understandingML} gives
\begin{equation}\label{eq:symmcost}
\E_{\mathcal O}\sup_{p\in\mathcal P}\{\Cbar{p}-\Chat(p)\}\le 2\mathcal R_N(\mathcal F').
\end{equation}
We bound \(R_N(\mathcal F')\) as follows. Write
\[
R_N(\mathcal F')
=
\E_{\mathcal O,\sigma}\!\left[
\sup_{p\in\mathcal P}
\frac{1}{N}\sum_{i=1}^N \sigma_i \frac{1}{n_i}\sum_{j=1}^{n_i} p(\bm X_{ij},\bm V_i)
\right].
\]
and pad each cluster to length \(n_{\max}\), with terms for \(j>n_i\) set to zero. Then
\[
\frac{1}{n_i}\sum_{j=1}^{n_i} p(\bm X_{ij},\bm V_i)
=
\sum_{j=1}^{n_{\max}} \frac{\bm{1}\{j\le n_i\}}{n_i}\, p(\bm X_{ij},\bm V_i),
\]
so
\[
R_N(\mathcal F')
=
\E_{\mathcal O,\sigma}\!\left[
\sup_{p\in\mathcal P}
\frac{1}{N}\sum_{i=1}^N \sigma_i
\sum_{j=1}^{n_{\max}}
\frac{\bm{1}\{j\le n_i\}}{n_i}\, p(\bm X_{ij},\bm V_i)
\right].
\]
Applying the triangle inequality,
\begin{align*}
R_N(\mathcal F')
\le &
\sum_{j=1}^{n_{\max}}
\E_{\mathcal O,\sigma}\!\left[
\sup_{p\in\mathcal P}
\frac{1}{N}\sum_{i=1}^N \sigma_i
\frac{\bm{1}\{j\le n_i\}}{n_i}\, p(\bm X_{ij},\bm V_i)
\right]\\
& \le \sum_{j = 1}^{n_{\max}}
\E_{\mathcal O,\sigma}\!\left[
\sup_{p\in\mathcal P}
\frac{1}{N}\sum_{i=1}^N \sigma_i p(\bm X_{ij},\bm V_i)
\right]\\
& =
n_{max}R_N(\mathcal P).
\end{align*}
since \(0\le \bm{1}\{j\le n_i\}/n_i \le 1\). Combining this calculation with \eqref{eq:symmcost} yields $\E_{\mathcal O}\sup_p\{\Cbar{p}-\Chat(p)\}\le 2n_{\max}\mathcal R_N(\mathcal P)$. By Rademacher symmetry, the same bound holds in the opposite direction. Write $Z_C^+=\sup_p\{\Cbar{p}-\Chat(p)\}$ and $Z_C^-=\sup_p\{\Chat(p)-\Cbar{p}\}$. Changing one cluster changes either supremum by at most $1/N$. Thus McDiarmid's inequality and a union bound give the stated high-probability result. Since $\mathcal P$ contains a constant policy, both $Z_C^+$ and $Z_C^-$ are nonnegative, and $\E_{\mathcal O}\max\{Z_C^+,Z_C^-\}\le\E_{\mathcal O}(Z_C^++Z_C^-)\le4n_{\max}\mathcal R_N(\mathcal P)$, proving the expected bound. 
\end{proof}

\begin{lemma}[Uniform deviation bounds for the average outcome]\label{lemma:outcome}
Suppose Assumptions \ref{ass:partial}--\ref{ass:regularity} hold. For any $\delta\in(0,1)$, with probability at least $1-\delta$,
\[
\sup_{p\in\mathcal P}|\Yhat{\ipw}(p)-\Ybar{p}|\le 2\sqrt2\frac{M}{\eta}n_{\max}^{3/2}\mathcal R_N(\mathcal P)+\frac{M}{\eta}\sqrt{\frac{2\log(2/\delta)}{N}}
\]
Moreover,
\[
\E_{\mathcal O}\left[\sup_{p\in\mathcal P}|\Yhat{\ipw}(p)-\Ybar{p}|\right]\le 2\sqrt2\frac{M}{\eta}n_{\max}^{3/2}\mathcal R_N(\mathcal P)+\frac{M}{\eta}\sqrt{\frac{2\log2}{N}}.
\]
\end{lemma}

\begin{proof}
Let $\psi^{\ipw}(\bm O_i,p)=n_i^{-1}\sum_{j=1}^{n_i}Y_{ij}(\bm A_i)\piclus{\bm A_i \mid \bm X_i, \bm V_i;p}/f(\bm A_i\mid\bm X_i,\bm V_i)$ and $\mathcal F''=\{\psi^{\ipw}(\cdot,p):p\in\mathcal P\}$. By identification, $\Ybar{p}=\E_{G_0}\psi^{\ipw}(\bm O_i,p)$ and $\Yhat{\ipw}(p)=N^{-1}\sum_i\psi^{\ipw}(\bm O_i,p)$. One-sided symmetrization gives
\begin{equation}\label{eq:symmoutcome}
\E_{\mathcal O}\sup_{p\in\mathcal P}\{\Ybar{p}-\Yhat{\ipw}(p)\}\le2\mathcal R_N(\mathcal F'').
\end{equation}
We now bound \(R_N(\mathcal F'')\). Write
\[
\psi^{\ipw}(\bm O_i,p)
=
W_i\,\piclus{\bm A_i \mid \bm X_i, \bm V_i;p},
\quad
W_i
=
\frac{1}{n_i}\sum_{j=1}^{n_i}
\frac{Y_{ij}(\bm A_i)}{f(\bm A_i\mid \bm X_i,\bm V_i)}.
\]
By bounded outcomes and overlap,
\(
|W_i|\le \frac{M}{\eta}.
\)
Hence, by contraction,
\[
R_N(\mathcal F'')
\le
\frac{M}{\eta}\,
\E_{\mathcal O,\sigma}\left[
\sup_{p\in\mathcal P}
\frac{1}{N}\sum_{i=1}^N \sigma_i \piclus{\bm A_i \mid \bm X_i, \bm V_i;p}
\right].
\]
Let
\[
\piclus{\bm A_i \mid \bm X_i, \bm V_i;p}
=
\prod_{j=1}^{n_i}
q_{ij}(p),
\quad
q_{ij}(p)
=
p(\bm X_{ij},\bm V_i)^{A_{ij}}
(1-p(\bm X_{ij},\bm V_i))^{1-A_{ij}}.
\]

The map \(u\mapsto \prod_{j=1}^{n_i} u_j\) is \(\sqrt{n_i}\)-Lipschitz w.r.t.\ \(\ell_2\), so by Corollary 4 in \citet{maurer_contraction},
\begin{align*}
R_N(\mathcal F'')
&\le
\frac{M}{\eta}\sqrt{2n_{\max}}\,
\E_{\mathcal O,\sigma}\left[
\sup_{p\in\mathcal P}
\frac{1}{N}\sum_{i=1}^N\sum_{j=1}^{n_i}
\sigma_{ij} q_{ij}(p)
\right] \\
&\le
\frac{M}{\eta}\sqrt{2n_{\max}}\,
\sum_{j=1}^{n_{\max}}
\E_{\mathcal O,\sigma}\left[
\sup_{p\in\mathcal P}
\frac{1}{N}\sum_{i=1}^N
\sigma_{ij}\bm{1}\{j\le n_i\} q_{ij}(p)
\right].
\end{align*}
For each $j=1,\ldots,n_i$, \(q_{ij}(p)\), is $1-$Lipschitz in $p$. Hence, by contraction:
\[
\E_{\mathcal O,\sigma}\left[
\sup_{p\in\mathcal P}
\frac{1}{N}\sum_{i=1}^N
\sigma_{ij}\bm{1}\{j\le n_i\} q_{ij}(p)
\right]
\le
R_N(\mathcal P).
\]
Therefore,
\[
R_N(\mathcal F'')
\le
\frac{M}{\eta}\sqrt{2n_{\max}}\cdot n_{\max} R_N(\mathcal P)
=
\frac{M}{\eta}\sqrt{2}\,n_{\max}^{3/2}R_N(\mathcal P).
\]
Combining with \eqref{eq:symmoutcome} and using Rademacher symmetry in the opposite direction, the one-sided deviations $Z_Y^\pm:=\sup_{p\in\mathcal P}\{\pm(\Yhat{\ipw}(p)-\Ybar{p})\}$ satisfy $\E_{\mathcal O}Z_Y^\pm\le A_{\ipw,N}$. Since $|\psi^{\ipw}(\bm O_i,p)|\le M/\eta$, replacing a cluster changes either $Z_Y^\pm$ by at most $2M/(N\eta)$. McDiarmid's inequality and a union bound give the stated high-probability inequality. Moreover, its exponential-moment form and the log-sum-exp inequality give
\begin{align*}
\E_{\mathcal O}\max\{Z_Y^+,Z_Y^-\}\le& 2\sqrt2\frac{M}{\eta}n_{\max}^{3/2}\mathcal R_N(\mathcal P)+\inf_{\lambda>0}\left\{\frac{\log2}{\lambda}+\frac{\lambda M^2}{2N\eta^2}\right\} \\
=&2\sqrt2\frac{M}{\eta}n_{\max}^{3/2}\mathcal R_N(\mathcal P)+\frac{M}{\eta}\sqrt{\frac{2\log2}{N}},
\end{align*}
which proves the expected bound.
\end{proof}


\begin{lemma}[Uniform deviation bounds for the H\'ajek average outcome]\label{lem:hajek_uniform}
Suppose Assumptions \ref{ass:partial}--\ref{ass:regularity} hold and $0<p(x,v)<1$ for all $p\in\mathcal P$ and $(x,v)\in\mathcal X\times\mathcal V$. For any $\delta\in(0,1)$, with probability at least $1-\delta$,
\[
\sup_{p\in\mathcal P}|\Yhat{\hj}(p)-\Ybar{p}|\le A_{\hj,N}+\frac{M}{\eta}\sqrt{\frac{2\log(2/\delta)}{N}},\qquad A_{\hj,N}:=6\sqrt2\frac{M}{\eta}n_{\max}^{3/2}\mathcal R_N(\mathcal P)+\frac{2M}{\eta\sqrt N}.
\]
Moreover,
\[
\E_{\mathcal O}\left[\sup_{p\in\mathcal P}|\Yhat{\hj}(p)-\Ybar{p}|\right]\le A_{\hj,N}+\frac{M}{\eta}\sqrt{\frac{2\log2}{N}}.
\]
\end{lemma}

\begin{proof}
Recall $w_i(p)=\piclus{\bm A_i \mid \bm X_i, \bm V_i;p}/f(\bm A_i\mid\bm X_i,\bm V_i)$ and $D_N(p)=N^{-1}\sum_i w_i(p)$. The strict positivity condition implies $w_i(p)>0$ and hence $D_N(p)>0$ for every $p\in\mathcal P$ and every $N$. Let us write 
\[
Ph=\E_{G_0}h(\bm O_i), \quad P_Nh=N^{-1}\sum_i h(\bm O_i).
\]
For $m\in[-M,M]$, let $g_{p,m}(\bm O_i)=w_i(p)(\Ybi-m)$ and $\mathcal G_H=\{g_{p,m}:p\in\mathcal P,\ m\in[-M,M]\}$. Since $Pw_p=1$ and $P(w_p\Ybi)=\Ybar{p}$, we have $Pg_{p,m}=\Ybar{p}-m$. Also, $\Yhat{\hj}(p)\in[-M,M]$ and $P_Ng_{p,\Yhat{\hj}(p)}=0$, so
\begin{equation}\label{eq:hajek-direct-identity}
\Yhat{\hj}(p)-\Ybar{p}=(P_N-P)g_{p,\Yhat{\hj}(p)}.
\end{equation}

Let $Z_H^+ = \sup_{g\in\mathcal G_H}(P_N-P)g$ and
$Z_H^- = \sup_{g\in\mathcal G_H}(P-P_N)g$ denote the
two one-sided uniform deviations, and let $Z_H=\max\{Z_H^+,Z_H^-\}$. We then have that, $\sup_p|\Yhat{\hj}(p)-\Ybar{p}|\le\max\{Z_H^+,Z_H^-\}$. By one-sided symmetrization and Rademacher symmetry, $\E_{\mathcal O}Z_H^\pm\le2\mathcal R_N(\mathcal G_H)$.

Let $\mathcal F_Y=\{\bm O_i\mapsto w_i(p)\Ybi:p\in\mathcal P\}$ and $\mathcal W=\{\bm O_i\mapsto w_i(p):p\in\mathcal P\}$. Writing $S_w(p)=N^{-1}\sum_i\sigma_iw_i(p)$, the supremum over $m\in[-M,M]$ gives $\mathcal R_N(\mathcal G_H)\le\mathcal R_N(\mathcal F_Y)+M\E_{\mathcal O,\sigma}\sup_p|S_w(p)|$. For any fixed $p_0\in\mathcal P$, Rademacher symmetry and $0\le w_i(p)\le1/\eta$ imply
\[
\E_{\mathcal O,\sigma}\sup_p|S_w(p)|\le2\mathcal R_N(\mathcal W)+\E_{\mathcal O,\sigma}|S_w(p_0)|\le2\mathcal R_N(\mathcal W)+\frac1{\eta\sqrt N}.
\]
The vector-contraction bound used in Lemma~\ref{lemma:outcome} gives $\mathcal R_N(\mathcal F_Y)\le\sqrt2(M/\eta)n_{\max}^{3/2}\mathcal R_N(\mathcal P)$ and $\mathcal R_N(\mathcal W)\le\sqrt2\eta^{-1}n_{\max}^{3/2}\mathcal R_N(\mathcal P)$. Therefore $\E_{\mathcal O}Z_H^\pm\le A_{\hj,N}$.

For each $p,m$, the range of $g_{p,m}$ has width at most $2M/\eta$. Thus changing one cluster alters either $Z_H^\pm$ by at most $2M/(N\eta)$. McDiarmid's inequality and a union bound give the high-probability statement. Finally, the exponential-moment bounded-differences inequality and the log-sum-exp bound imply
\[
\E_{\mathcal O}\max\{Z_H^+,Z_H^-\}\le A_{\hj,N}+\inf_{\lambda>0}\left\{\frac{\log2}{\lambda}+\frac{\lambda M^2}{2N\eta^2}\right\}=A_{\hj,N}+\frac{M}{\eta}\sqrt{\frac{2\log2}{N}},
\]
which proves the expected bound.
\end{proof}

\begin{lemma}[Margin]\label{lemma:margin}
Let $C_0 > 1$, let $\widetilde Y : \mathcal P \to \mathbb R$
be any estimator of $\bar Y$, let $m : \mathcal P \to (0,\infty)$
be any, possibly data-dependent, margin, and let
\[
\hat p \in \arg\min_{p\in\mathcal P}
\left\{
\Chat(p) + C_0 h_{m(p)}(\widetilde Y(p);T)
\right\}.
\]
If $q\in\mathcal P$ satisfies
$\widetilde Y(q)\ge T+m(q)$, then (a) $\Chat(\hat p)\leq\Chat(q)$; (b) $\widetilde Y(\hat p)\ge T+c\,m(\hat p)$, where $c=c(C_0)$ is defined in Section 4.2.
\end{lemma}

\begin{proof}
Since $\widetilde Y(q)\ge T+m(q)$,
we have $h_{m(q)}(\widetilde Y(q);T)=0$.
By optimality of $\hat p$,
\[
\Chat(\hat p)
+
C_0 h_{m(\hat p)}(\widetilde Y(\hat p);T)
\leq
\Chat(q).
\]
Since the penalty is nonnegative, (a) follows.
Moreover, since $0\leq \Chat(p)\leq 1$,
\[
h_{m(\hat p)}(\widetilde Y(\hat p);T)
\leq \frac{1}{C_0}<1.
\]
We now consider two cases.
If $1<C_0\leq 2$, then $1/C_0\geq 1/2$.
By the definition of $h_\epsilon$, if
$\widetilde Y(\hat p)<T+m(\hat p)/2$, then
\[
1-\frac{2}{m(\hat p)^2}
\bigl(\widetilde Y(\hat p)-T\bigr)^2
\leq \frac{1}{C_0},
\]
which implies
\[
\widetilde Y(\hat p)
\geq T+m(\hat p)
\sqrt{\frac{1-1/C_0}{2}}.
\]
The same inequality holds if
$\widetilde Y(\hat p)\geq T+m(\hat p)/2$,
since $\sqrt{(1-1/C_0)/2}\leq 1/2$.

If $C_0\geq 2$, then $1/C_0\leq 1/2$,
which implies
$\widetilde Y(\hat p)\geq T+m(\hat p)/2$.
If $\widetilde Y(\hat p)<T+m(\hat p)$,
the definition of $h_\epsilon$ gives
\[
\frac{2}{m(\hat p)^2}
\bigl(T+m(\hat p)-\widetilde Y(\hat p)\bigr)^2
\leq \frac{1}{C_0},
\]
and therefore
\[
\widetilde Y(\hat p)
\geq T+m(\hat p)
\left(1-\frac{1}{\sqrt{2C_0}}\right).
\]
This inequality also holds when
$\widetilde Y(\hat p)\geq T+m(\hat p)$.
Thus, in both cases,
\(
\widetilde Y(\hat p)\geq T+c\,m(\hat p),
\)
establishing (b).
\end{proof}

\paragraph{Bernstein preliminaries.}
For a class $\mathcal F$ and $r>0$, write $N_\infty(r,\mathcal F,k)=\sup_{\bm o\in\mathcal O^k}N(r,\mathcal F(\bm o),\|\cdot\|_\infty)$, where the covering centers belong to $\mathcal F(\bm o)$. Let $Pf=\E_{G_0}f(O)$, $P_Nf=N^{-1}\sum_i f(O_i)$ and $s_N(f)^2=(N-1)^{-1}\sum_i\{f(O_i)-P_Nf\}^2$. All the Bernstein results below require $N\ge16$.

\begin{theorem}[Uniform empirical Bernstein bounds]\label{th:bernsteinbounds}
Suppose Assumptions~\ref{ass:partial}--\ref{ass:regularity} hold and $N\ge16$. Set $B_\psi=M/\eta$, $\Psi=\{\psi^{\ipw}(\cdot,p):p\in\mathcal P\}$ and $\widetilde\Psi=\{(\psi+B_\psi)/(2B_\psi):\psi\in\Psi\}$. Define the common complexity term
\[
\Lambda_{N,\ipw}(\delta)=\log\{20N_\infty(1/N,\widetilde\Psi,2N)/\delta\}.
\]
With probability at least $1-\delta$, simultaneously for all $p\in\mathcal P$,
\[
|\Yhat{\ipw}(p)-\Ybar{p}|
\le \sqrt{18\,s_N(\psi_p^{\ipw})^2\Lambda_{N,\ipw}(\delta)/N}
+32B_\psi\Lambda_{N,\ipw}(\delta)/N.
\]
Moreover, 
\begin{equation}\label{eq:inequalitycovering}
\Lambda_{N,\ipw}(\delta)\le\log\{20N_\infty(2/(Nn_{\max}),\mathcal P,2Nn_{\max})/\delta\}.
\end{equation}
\end{theorem}
\begin{proof}
Bounded outcomes and overlap give $|\psi_p^{\ipw}|\le B_\psi$, so $\widetilde\psi_p=(\psi_p^{\ipw}+B_\psi)/(2B_\psi)\in[0,1]$. Apply Theorem~6 of \citet{maurerpontil} separately to $\widetilde\Psi$ and $1-\widetilde\Psi$, each at level $\delta/2$. Their covering numbers and empirical variances agree. A union bound therefore gives, uniformly in $p$,
\[
|(P-P_N)\widetilde\psi_p|
\le\sqrt{18\,s_N(\widetilde\psi_p)^2\Lambda_{N,\ipw}(\delta)/N}
+15\Lambda_{N,\ipw}(\delta)/(N-1).
\]
Multiplying by $2B_\psi$ and using $30/(N-1)\le32/N$ for $N\ge16$ proves the deviation bound.

For the covering bound, fix $2N$ clusters and an $r$-cover of $\mathcal P$ on their at most $2Nn_{\max}$ covariate pairs. The telescoping product inequality gives $|\piclus{\bm A_i \mid \bm X_i, \bm V_i;p}-\piclus{\bm A_i \mid \bm X_i, \bm V_i;q}|\le n_{\max}r$ and hence $|\widetilde\psi_p(O_i)-\widetilde\psi_q(O_i)|\le n_{\max}r/2$. Taking $r=2/(Nn_{\max})$ proves the claim.
\end{proof}

\begin{lemma}[Uniform bounds of empirical standard deviations]\label{lemma:variancebounds}
Under the assumptions of Theorem~\ref{th:bernsteinbounds}, let $\sigma_{\ipw}(p)^2=\Varop_{G_0}\{\psi_p^{\ipw}(O)\}$ and $\sdhat{\ipw}(p)=s_N(\psi_p^{\ipw})$. With probability at least $1-\delta$,
\[
\sup_{p\in\mathcal P}|\sdhat{\ipw}(p)-\sigma_{\ipw}(p)|
\le16B_\psi\sqrt{\Lambda_{N,\ipw}(\delta)/N}.
\]
\end{lemma}
\begin{proof}
We give the uniform argument for any class $\mathcal F\subset[0,1]$. Set $K=N_\infty(1/N,\mathcal F,2N)$ and $\kappa_N=\sqrt{N/(N-1)}$. For vectors $u,v$, centering is a contraction in Euclidean norm, so $|s_N(u)-s_N(v)|\le\kappa_N\|u-v\|_\infty$.

Let $O'$ be an independent sample and $\sigma(f)=\sqrt{\Varop\{f(O)\}}$. By both tails of Theorem~10 of \citet{maurerpontil}, for any fixed $f$ the event $|s_N(f;O')-\sigma(f)|\le t/2$ has probability at least $1/2$ whenever $t/2\ge\sqrt{2\log4/(N-1)}$. Choosing a function witnessing a deviation (or an arbitrarily close witness), and conditioning on $O$, gives
\[
\Pr\{\sup_f|s_N(f;O)-\sigma(f)|>t\}
\le2\Pr\{\sup_f|s_N(f;O)-s_N(f;O')|>t/2\}.
\]
Condition on the paired observations and swap each pair independently with probability $1/2$. For each function, the two resulting vectors have independent coordinates and identical marginal distributions; their expected sample variances are equal. Theorem~10 and a union bound over its two tails for both vectors give $\Pr_{\rm swap}\{|s_N(f;O)-s_N(f;O')|>u\}\le4\exp\{-(N-1)u^2/8\}$. A $1/N$-cover on the combined sample has at most $K$ centers, and approximation changes the difference of standard deviations by at most $2\kappa_N/N$. Thus, for $t/2>2\kappa_N/N$,
\[
\Pr\{\sup_f|s_N(f;O)-\sigma(f)|>t\}
\le8K\exp\{-(N-1)(t/2-2\kappa_N/N)^2/8\}.
\]
Take $t=4\kappa_N/N+\sqrt{32\log(8K/\delta)/(N-1)}$. This satisfies the preceding conditions and gives failure probability at most $\delta$. For $N\ge16$, $t\le8\sqrt{\log(20K/\delta)/N}$. Apply this result to $\widetilde\Psi$ and multiply by $2B_\psi$. As usual, the suprema are understood to be measurable.
\end{proof}

\paragraph{H\'ajek Bernstein preliminaries.}
Assume throughout these H\'ajek results that $0<p(\bm x,\bm v)<1$ for every $p\in\mathcal P$ and covariate pair $(x,v)$. As in Lemma~\ref{lem:hajek_uniform}, this implies $w_i(p)>0$ and $D_N(p)>0$ for every sample. Reuse its class $\mathcal G_H=\{g_{p,m}:p\in\mathcal P,\ m\in[-M,M]\}$, where $g_{p,m}(O_i)=w_i(p)(\Ybi-m)$, and its identity~\eqref{eq:hajek-direct-identity}.

Define $\phihj(p)=g_{p,\Ybar{p}}(O_i)$ and $\sigma_{\hj}(p)^2=\Varop_{G_0}\{\phihj(p)\}$. Let $\widetilde\phihj(p)=g_{p,\Yhat{\hj}(p)}(O_i)$. Its empirical mean is zero, and
\begin{equation}\label{eq:H-residual-var}
\widetilde\sigma_{H,N}(p)^2=s_N(\widetilde\phi^{\hj}(p))^2
=\frac1{N-1}\sum_i w_i(p)^2\{\Ybi-\Yhat{\hj}(p)\}^2.
\end{equation}
Thus $\widetilde\sigma_{H,N}(p)=D_N(p)\sdhat{\hj}(p)$. Also set $\widehat\sigma_w(p)=s_N(w_p)$ and $B_{\phi,H}=2M/\eta$. Each $g_{p,m}$ has range contained in $[-(M+m)/\eta,(M-m)/\eta]$, of width $B_{\phi,H}$. Its normalized version is $\widetilde g_{p,m}=g_{p,m}/B_{\phi,H}+(M+m)/(2M)\in[0,1]$; write $\widetilde\Phi_H^\star=\{\widetilde g_{p,m}:p\in\mathcal P,\ m\in[-M,M]\}$.

\begin{theorem}[Uniform empirical Bernstein bound for H\'ajek]\label{prop:hajek-bernstein}
Suppose Assumptions~\ref{ass:partial}--\ref{ass:regularity} hold, $N\ge16$, and the strict policy positivity condition above holds. Define
\[
\Lambda_{N,\hj}(\delta)=\log\{20N_\infty(1/N,\widetilde\Phi_H^\star,2N)/\delta\}.
\]
Set $a_{N,H}(\delta)=\sqrt{18\Lambda_{N,\hj}(\delta)/N}$ and $b_{N,H}(\delta)=16B_{\phi,H}\Lambda_{N,\hj}(\delta)/N$. With probability at least $1-\delta$, simultaneously for all $p\in\mathcal P$,
\begin{align}
|\Yhat{\hj}(p)-\Ybar{p}|&\le\alphahat{\hj}(p),\label{eq:H-EB-exact}\\
\alphahat{\hj}(p)&=a_{N,H}(\delta)\widetilde\sigma_{H,N}(p)+b_{N,H}(\delta).\label{eq:H-alpha-exact}
\end{align}
Moreover,
\begin{equation}\label{eq:H-cover}
N_\infty(1/N,\widetilde\Phi_H^\star,2N)
\le(N+1)N_\infty(1/(2Nn_{\max}),\mathcal P,2Nn_{\max}).
\end{equation}
\end{theorem}
\begin{proof}
Apply Theorem~6 of \citet{maurerpontil} to $\widetilde\Phi_H^\star$ and $1-\widetilde\Phi_H^\star$, each at level $\delta/2$. The shift in $\widetilde g_{p,m}$ is constant across observations, so it affects neither deviations nor standard deviations. Rescaling by $B_{\phi,H}$ and using $15/(N-1)\le16/N$ gives, uniformly in $(p,m)$,
\begin{equation}\label{eq:H-EB-joint}
|(P-P_N)g_{p,m}|\le a_{N,H}(\delta)s_N(g_{p,m})+b_{N,H}(\delta).
\end{equation}
Evaluate this uniform inequality at $m=\Yhat{\hj}(p)\in[-M,M]$. Identity~\eqref{eq:hajek-direct-identity} and \eqref{eq:H-residual-var} give the claimed bound.

For the covering bound, an $r$-cover of $\mathcal P$ on the covariates of $2N$ clusters gives $|w_i(p)-w_i(q)|\le n_{\max}r/\eta$. Since $0\le\eta w_i(q)\le1$,
\[
|\widetilde g_{p,m}(O_i)-\widetilde g_{q,m'}(O_i)|
\le n_{\max}r+|m-m'|/(2M).
\]
Take $r=1/(2Nn_{\max})$ and an $M/N$-cover of $[-M,M]$ with at most $N+1$ centers. Their product is the required $1/N$-cover.
\end{proof}

\begin{lemma}[Control of the H\'ajek variance-adaptive margin]\label{lem:hajek-margin}
Under the assumptions of Theorem~\ref{prop:hajek-bernstein}, with probability at least $1-\delta$,
\begin{equation}\label{eq:H-sd-pop}
\sup_{p\in\mathcal P}|s_N(\phi^{\hj}(p))-\sigma_{\hj}(p)|
\le8B_{\phi,H}\sqrt{\Lambda_{N,\hj}(\delta)/N}.
\end{equation}
If, additionally,
\begin{equation}\label{eq:H-weight-condition}
a_{N,H}(\delta)\sup_{p\in\mathcal P}\widehat\sigma_w(p)\le1/2,
\end{equation}
then, on the intersection of this event and the event of Theorem~\ref{prop:hajek-bernstein}, uniformly over $p$,
\[
\alphahat{\hj}(p)\le2\sqrt{18}\,\sigma_{\hj}(p)\sqrt{\Lambda_{N,\hj}(\delta)/N}
+(16\sqrt{18}+32)B_{\phi,H}\Lambda_{N,\hj}(\delta)/N.
\]
\end{lemma}
\begin{proof}
Apply the normalized-class argument in the proof of Lemma~\ref{lemma:variancebounds} to $\widetilde\Phi_H^\star$. Standard deviations are invariant under the $m$-dependent constant shift and scale by $B_{\phi,H}$. Since $\Ybar{p}\in[-M,M]$, the population-centered score belongs to $\mathcal G_H$, giving \eqref{eq:H-sd-pop}.

Write $a=a_{N,H}(\delta)$, $b=b_{N,H}(\delta)$ and $\delta_p=\Yhat{\hj}(p)-\Ybar{p}$. The sample standard deviation is a seminorm, so $\widetilde\sigma_{H,N}(p)\le s_N(\phi^{\hj}(p))+|\delta_p|\widehat\sigma_w(p)$. On the Bernstein event, $|\delta_p|\le\alphahat{\hj}(p)$, and hence $\{1-a\widehat\sigma_w(p)\}\alphahat{\hj}(p)\le a\,s_N(\phi^{\hj}(p))+b$. Condition~\eqref{eq:H-weight-condition} gives $\alphahat{\hj}(p)\le2a\,s_N(\phi^{\hj}(p))+2b$. Substitute \eqref{eq:H-sd-pop} to conclude.
\end{proof}

\subsection{Proof of Theorem \ref{th:regret}}
\begin{proof}[Proof]
Recall the notation \(R_{G_0}^\star = \inf_{p \in \mathcal{P}} R_{G_0}(p)\) and 
\(
V_{\epsilon,G_0}^\star = \inf_{p \in \mathcal{P}} V_{\epsilon, G_0}(p)
\).
Consider the decomposition
\begin{align}
R_{G_0}(\hat p_{\ipw})-R_{G_0}^\star
={}&
\bigl(R_{G_0}(\hat p_{\ipw})-V_{\epsilon,G_0}(\hat p_{\ipw})\bigr)
+\bigl(V_{\epsilon,G_0}(\hat p_{\ipw})-V_{\epsilon,G_0}^\star\bigr) \notag\\
&+\bigl(V_{\epsilon,G_0}^\star-R_{G_0}^\star\bigr).
\label{eq:regret-decomp}
\end{align}
Since \(V_{\epsilon,G_0}(p)\ge R_{G_0}(p)\) for every \(p\in\mathcal P\), the first term on the right-hand side is non-positive. Therefore,
\begin{align}
R_{G_0}(\hat p_{\ipw})-R_{G_0}^\star
\le
\underbrace{V_{\epsilon,G_0}^\star-R_{G_0}^\star}_{\text{approximation error}}
+
\underbrace{V_{\epsilon,G_0}(\hat p_{\ipw})-V_{\epsilon,G_0}^\star}_{\text{surrogate estimation error}}.
\label{eq:regret-decomp-2}
\end{align}
By empirical optimality of $\phat{\ipw}$ and the
$2/\epsilon$-Lipschitz property of $h_\epsilon$,
\[
V_{\epsilon,G_0}(\phat{\ipw})-V_{\epsilon,G_0}^\star
\le 2\sup_{p\in\mathcal P}|V_{\epsilon,G_0}(p)-\Vhat{\epsilon,\ipw}(p)|
\le 2\sup_{p\in\mathcal P}|\Cbar{p}-\Chat(p)|
+\frac{4C_0}{\epsilon}\sup_{p\in\mathcal P}|\Ybar{p}-\Yhat{\ipw}(p)|.
\]
Taking expectations and applying Lemmas~\ref{lemma:cost} and
\ref{lemma:outcome} gives
\[
\E_{\mathcal O}[V_{\epsilon,G_0}(\phat{\ipw})-V_{\epsilon,G_0}^\star]
\le 8n_{\max}\mathcal R_N(\mathcal P)
+\frac{8\sqrt2 C_0M}{\eta\epsilon}n_{\max}^{3/2}\mathcal R_N(\mathcal P)
+\frac{4C_0M}{\eta\epsilon}\sqrt{\frac{2\log2}{N}}.
\]
Combining this with \eqref{eq:regret-decomp-2} proves the theorem.
\end{proof}

\subsection{Proof of Theorem \ref{th:approxerr}}
\begin{proof}[Proof]
Recall the following
\[
R_{G_0}(p)=\Cbar{p}+C_0\,\bm{1}\{\Ybar{p}<T\}, \qquad V_{\epsilon,G_0}(p)=\Cbar{p}+C_0\,h_\epsilon(\Ybar{p};T),
\]
Note that \(p_\epsilon^\star\) minimizes \(V_{\epsilon,G_0}\) over \(\mathcal P\). Since $p^\star$ solves the original constrained problem, \(\bar Y(p^\star)\geq T\).
For sufficiently small \(\epsilon>0\), we have \(\epsilon/\kappa \le t_0\). Set
\(
\tilde p = p^\star + \frac{\epsilon}{\kappa} d.
\)
By the growth condition,
\[
\bar Y(\tilde p)
\ge
\bar Y(p^\star)+\kappa\frac{\epsilon}{\kappa}
\ge
T+\epsilon.
\]
Hence \(h_\epsilon(\bar Y(\tilde p);T)=0\), so
\(
V_{\epsilon,G_0}(\tilde p)=\bar C(\tilde p).
\), and by optimality of \(p_\epsilon^\star\),
\(
V_{\epsilon,G_0}^\star=V_{\epsilon,G_0}(p_\epsilon^\star)\le V_{\epsilon,G_0}(\tilde p)=\bar C(\tilde p).
\)
On the other hand, 
we have
$
R_{G_0}^\star=R_{G_0}(p^\star)=\bar C(p^\star),
$. Therefore,
\[
V_{\epsilon,G_0}^\star-R_{G_0}^\star
\le
\bar C_{G_0}(\tilde p)-\bar C_{G_0}(p^\star).
\]
By Lipschitz continuity of \(\bar C_{G_0}\),
\[
\bar C_{G_0}(\tilde p)-\bar C_{G_0}(p^\star)
\le
L_p\|\tilde p-p^\star\|
=
\frac{L_p\|d\|}{\kappa}\epsilon
\implies
V_{\epsilon,G_0}^\star-R_{G_0}^\star\le \frac{L_p\|d\|}{\kappa}\epsilon.
\]

\end{proof}

\subsection{Proof of Theorem \ref{th:raterefinement}}\label{appendix:th3}
\begin{proof}
Define the following set 
\[
\mathcal P_N^+
=
\left\{
p\in\mathcal P:
\bar Y(p)\ge T+\epsilon_N+\Bdev{\ipw}(\delta_N)
\right\}.
\]
Since $\epsilon_N+\Bdev{\ipw}(\delta_N)<s$,
this set is nonempty. Further define the event 
\[
E_Y=
\left\{
\sup_{p\in\mathcal P}
\left|\Yhat{\ipw}(p)-\bar Y(p)\right|
\le \Bdev{\ipw}(\delta_N)
\right\}.
\]
By Lemma \ref{lemma:outcome},
$\Pr(E_Y^c)\le\delta_N$, where $E_Y^c$ defines the complement event of $E_Y$.
On $E_Y$, every $p\in\mathcal P_N^+$ satisfies
\[
\Yhat{\ipw}(p)
\ge \bar Y(p)-\Bdev{\ipw}(\delta_N)
\ge T+\epsilon_N.
\]
Therefore, applying Lemma \ref{lemma:margin}
with constant margin $m:=\epsilon_N$ yields
\[
\Chat(\hat p_{\ipw})
\le \inf_{p\in\mathcal P_N^+}\Chat(p),
\qquad
\Yhat{\ipw}(\hat p_{\ipw})\ge T+c\epsilon_N.
\]
Since $c\epsilon_N\ge\Bdev{\ipw}(\delta_N)$,
on $E_Y$ we also have
\[
\bar Y(\hat p_{\ipw})
\ge \Yhat{\ipw}(\hat p_{\ipw})
-\Bdev{\ipw}(\delta_N)
\ge T+c\epsilon_N-\Bdev{\ipw}(\delta_N)
\ge T.
\]
Hence, the estimated policy satisfies the population
outcome constraint on $E_Y$.

Moreover, on $E_Y$,
\[
\Cbar{\hat p_{\ipw}}
\le \inf_{p\in\mathcal P_N^+}\Cbar{p}
+2\sup_{p\in\mathcal P}|\Chat(p)-\Cbar{p}|.
\]
Since $\Ybar{\hat p_{\ipw}}\ge T$ on $E_Y$, it follows that
\[
R_{G_0}(\hat p_{\ipw})-R_{G_0}^\star
\le \bar C^\star(T+\epsilon_N+\Bdev{\ipw}(\delta_N))
-\bar C^\star(T)
+2\sup_{p\in\mathcal P}|\Chat(p)-\Cbar{p}|.
\]
On $E_Y^c$, regret is bounded by $1+C_0$. Taking expectations and applying Lemma~\ref{lemma:cost} therefore gives
\[
\Reg(\hat p_{\ipw})
\le \bar C^\star(T+\epsilon_N+\Bdev{\ipw}(\delta_N))
-\bar C^\star(T)
+8n_{\max}\mathcal R_N(\mathcal P)
+(1+C_0)\delta_N.
\]
\end{proof}


\subsection{Proof of Theorem \ref{th:Hajek1}}\label{appendix:thHajek1}

\begin{proof}
For the first statement, the regret decomposition used in Theorem \ref{th:regret} and empirical optimality of $\phat{\hj}$ give
\begin{align*}
R_{G_0}(\phat{\hj})-R_{G_0}^\star
&\le V_{\epsilon,G_0}^\star-R_{G_0}^\star
+2\sup_{p\in\mathcal P}|V_{\epsilon,G_0}(p)-\Vhat{\epsilon,\hj}(p)|\\
&\le V_{\epsilon,G_0}^\star-R_{G_0}^\star
+2\sup_{p\in\mathcal P}|\Chat(p)-\Cbar{p}|
+\frac{4C_0}{\epsilon}\sup_{p\in\mathcal P}|\Yhat{\hj}(p)-\Ybar{p}|,
\end{align*}
where the second inequality uses the $2/\epsilon$-Lipschitz property of $h_\epsilon$. Taking expectations and applying Lemmas~\ref{lemma:cost} and \ref{lem:hajek_uniform} gives
\[
\Reg(\phat{\hj})
\le V_{\epsilon,G_0}^\star-R_{G_0}^\star
+8n_{\max}\mathcal R_N(\mathcal P)
+24\sqrt2\frac{C_0M}{\eta\epsilon}n_{\max}^{3/2}\mathcal R_N(\mathcal P)
+\frac{4C_0M}{\eta\epsilon\sqrt N}(2+\sqrt{2\log2}),
\]
proving the first statement.

For the second statement, let $B_H=\Bdev{\hj}(\delta_N/2)$ and define $\mathcal P_N^+=\{p\in\mathcal P:\Ybar{p}\ge T+\epsilon_N+B_H\}$ and $E_H=\{\sup_{p\in\mathcal P}|\Yhat{\hj}(p)-\Ybar{p}|\le B_H\}$. By Lemma~\ref{lem:hajek_uniform}, $\Pr(E_H^c)\le\delta_N/2$. Moreover, $\mathcal P_N^+$ is nonempty since $\epsilon_N+B_H<s$.

On $E_H$, every $p\in\mathcal P_N^+$ satisfies $\Yhat{\hj}(p)\ge T+\epsilon_N$. Thus, Lemma~\ref{lemma:margin} gives $\Chat(\phat{\hj})\le\inf_{p\in\mathcal P_N^+}\Chat(p)$ and $\Yhat{\hj}(\phat{\hj})\ge T+c\epsilon_N$. Since $c\epsilon_N\ge B_H$, we also have $\Ybar{\phat{\hj}}\ge T$ on $E_H$.

Furthermore, on $E_H$,
\[
\Cbar{\phat{\hj}}
\le \inf_{p\in\mathcal P_N^+}\Cbar{p}
+2\sup_{p\in\mathcal P}|\Chat(p)-\Cbar{p}|
=\bar C^\star(T+\epsilon_N+B_H)
+2\sup_{p\in\mathcal P}|\Chat(p)-\Cbar{p}|.
\]
Since $R_{G_0}^\star=\bar C^\star(T)$ and regret is at most $1+C_0$ on $E_H^c$, taking expectations and applying Lemma~\ref{lemma:cost} yields
\[
\Reg(\phat{\hj})
\le \bar C^\star(T+\epsilon_N+B_H)-\bar C^\star(T)
+8n_{\max}\mathcal R_N(\mathcal P)
+(1+C_0)\delta_N.
\]
Finally, if $\mathcal R_N(\mathcal P)=O(N^{-1/2})$, $\delta_N=N^{-1}$ and $\epsilon_N=B_H/c$, then $B_H=O(\sqrt{\log N/N})$ and $\epsilon_N+B_H\to0$. Under the assumed local behavior of $\bar C^\star$, the preceding bound is $O(\sqrt{\log N/N})$, proving the second statement.
\end{proof}

\subsection{Proof of Theorem \ref{th:regretboundVA}}\label{app:proofVA}
\begin{proof}
Write $L=\Lambda_N=\Lambda_{N,\ipw}(\delta_N/2)$ and $B_\psi=M/\eta$. Since $\sigma_{\ipw}(p)\le B_\psi$ uniformly and $L/N\to0$, $\sup_p\Delta_N(p)\to0$. The positive slack assumption supplies a fixed policy with $\Ybar{p}>T$, so $\mathcal A_N^\Delta$ is nonempty for all sufficiently large $N$. Increase $N_0$ if necessary so that $N_0\ge16$.

Let $\hat p=\phat{\ipw}^{\mathrm{VA}}$, and, for any $\xi>0$, choose $q\in\mathcal A_N^\Delta$ with $\Cbar{q}\le\inf_{p\in\mathcal A_N^\Delta}\Cbar{p}+\xi$. Define
\[
\mathcal E_Y=\{|\Yhat{\ipw}(p)-\Ybar{p}|\le c\widehat\epsilon_N(p)\ \forall p\},\qquad
\mathcal E_\epsilon=\{(1+c)\widehat\epsilon_N(p)\le\Delta_N(p)\ \forall p\}.
\]
Theorem~\ref{th:bernsteinbounds} gives $\Pr(\mathcal E_Y^c)\le\delta_N/2$. By Lemma~\ref{lemma:variancebounds}, with probability at least $1-\delta_N/2$, $c\widehat\epsilon_N(p)\le\sqrt{18}\,\sigma_{\ipw}(p)\sqrt{L/N}+(16\sqrt{18}+32)B_\psi L/N$ uniformly in $p$. Since $d_1\ge(1+c^{-1})\sqrt{18}$ and $d_2\ge(1+c^{-1})(16\sqrt{18}+32)$, the event $\mathcal E=\mathcal E_Y\cap\mathcal E_\epsilon$ has probability at least $1-\delta_N$.

On $\mathcal E$, $\Yhat{\ipw}(q)\ge\Ybar{q}-c\widehat\epsilon_N(q)\ge T+\widehat\epsilon_N(q)$. Lemma~\ref{lemma:margin}, applied with $m(p)=\widehat\epsilon_N(p)$, gives $\Chat(\hat p)\le\Chat(q)$ and $\Yhat{\ipw}(\hat p)\ge T+c\widehat\epsilon_N(\hat p)$. Thus $\Ybar{\hat p}\ge\Yhat{\ipw}(\hat p)-c\widehat\epsilon_N(\hat p)\ge T$ on $\mathcal E$, and $\Pr\{\Ybar{\hat p}<T\}\le\delta_N$.

Let $Z_C^\pm=\sup_p\{\pm(\Chat(p)-\Cbar{p})\}$. On $\mathcal E$, $\Cbar{\hat p}-\Cbar{q}\le Z_C^++Z_C^-$. On $\mathcal E^c$, the cost difference is at most one. The sum $Z_C^++Z_C^-$ is always nonnegative; the two one-sided bounds in the proof of Lemma~\ref{lemma:cost} give $\E_{\mathcal O}(Z_C^++Z_C^-)\le4n_{\max}\mathcal R_N(\mathcal P)$. Consequently,
\[
\E_{\mathcal O}\{\Cbar{\hat p}-\Cbar{q}\}\le4n_{\max}\mathcal R_N(\mathcal P)+\delta_N.
\]
As in Theorem~\ref{th:raterefinement}, nonempty population feasibility gives $R_{G_0}^\star=\bar C^\star(T)$. Add the constraint penalty $C_0\Pr\{\Ybar{\hat p}<T\}$ to the preceding bound and let $\xi\downarrow0$. This proves the displayed regret inequality. The final rate follows from the additional conditions in the theorem.
\end{proof}

\subsection{Proof of Theorem \ref{th:hajekva}}\label{app:hajek-bernstein}
Use the preliminary results above and evaluate $\alphahat{\hj}$ at $\delta=\delta_N/2$. For $d_1,d_2$ depending only on $C_0$, define
\begin{equation}\label{eq:hajek-pop-buffer}
\Delta_{N,\hj}(p)=d_1\sigma_{\hj}(p)\sqrt{\Lambda_{N,\hj}(\delta_N/2)/N}
+d_2B_{\phi,H}\Lambda_{N,\hj}(\delta_N/2)/N.
\end{equation}
Let $\mathcal A_{N,\hj}^{\Delta}=\{p\in\mathcal P:\Ybar{p}\ge T+\Delta_{N,\hj}(p)\}$ and, when this set is nonempty, $\bar C^{\star,\mathrm{VA}}_{N,\hj}=\inf_{p\in\mathcal A_{N,\hj}^{\Delta}}\Cbar{p}$.

Since $0\le w_i(p)\le1/\eta$, $\sup_p\widehat\sigma_w(p)\le\eta^{-1}\sqrt{N/(N-1)}$. Hence \eqref{eq:H-weight-condition}, evaluated at $\delta_N/2$, is automatic for all sufficiently large $N$ if $\Lambda_{N,\hj}(\delta_N/2)=o(N)$. Take $d_1\ge(1+c^{-1})2\sqrt{18}$ and $d_2\ge(1+c^{-1})(16\sqrt{18}+32)$. Theorem~\ref{prop:hajek-bernstein} and Lemma~\ref{lem:hajek-margin}, each at level $\delta_N/2$, then imply, with probability at least $1-\delta_N$,
\begin{equation}\label{eq:H-two-events}
|\Yhat{\hj}(p)-\Ybar{p}|\le\alphahat{\hj}(p),\qquad
(1+c^{-1})\alphahat{\hj}(p)\le\Delta_{N,\hj}(p)\quad\text{for all }p\in\mathcal P.
\end{equation}
Here positivity follows directly from the strict policy positivity condition of Lemma~\ref{lem:hajek_uniform}.

\begin{theorem}[Variance-adaptive H\'ajek refinement]\label{th:hajekva}
Suppose Assumptions~\ref{ass:partial}--\ref{ass:regularity} hold, $N\ge16$, $\delta_N\to0$, and $0<p(x,v)<1$ for all $p\in\mathcal P$ and covariate pairs $(x,v)$. Suppose also that \eqref{eq:H-weight-condition} holds at level $\delta_N/2$ for every sample and $\mathcal A_{N,\hj}^{\Delta}\neq\varnothing$. Set $\widehat\epsilon_{N,\hj}(p)=\alphahat{\hj}(p)/c$, evaluated at $\delta=\delta_N/2$, and let
\[
\phat{\hj}^{\mathrm{VA}}\in\arg\min_{p\in\mathcal P}
\{\Chat(p)+C_0h_{\widehat\epsilon_{N,\hj}(p)}(\Yhat{\hj}(p);T)\}.
\]
Then
\begin{equation}\label{eq:boundhjva}
\E_{\mathcal O}\{R_{G_0}(\phat{\hj}^{\mathrm{VA}})-R_{G_0}^\star\}
\le\bar C^{\star,\mathrm{VA}}_{N,\hj}-\bar C^\star(T)
+4n_{\max}\mathcal R_N(\mathcal P)+(1+C_0)\delta_N.
\end{equation}
\end{theorem}
\begin{proof}
By \eqref{eq:H-two-events}, the two uniform events used in the proof of Theorem~\ref{th:regretboundVA} hold for the H\'ajek estimator with probability at least $1-\delta_N$, using $\widehat\epsilon_{N,\hj}(p)$ in place of $\widehat\epsilon_N(p)$ and $\Delta_{N,\hj}$ in place of $\Delta_N$. Apply the same argument with $\Yhat{\hj}$ in place of $\Yhat{\ipw}$. In particular, every $q\in\mathcal A_{N,\hj}^{\Delta}$ has zero empirical penalty on this event; Lemma~\ref{lemma:margin} gives $\Chat(\hat p)\le\Chat(q)$ and $\Yhat{\hj}(\hat p)\ge T+\alphahat{\hj}(\hat p)$, hence population feasibility. The cost argument is unchanged and gives $4n_{\max}\mathcal R_N(\mathcal P)+\delta_N$; the constraint penalty contributes $C_0\delta_N$. Taking the infimum over comparison policies proves the result.
\end{proof}

\subsection{Proof of Theorem \ref{thm:dual_equiv}}

\begin{proof}
The functional $\bar C$ is $1$-Lipschitz in the sup norm:
$|\bar C(p')-\bar C(p)|\le\lVert p'-p\rVert_\infty$.
Suppose a budget-constrained maximizer $p^\star$ satisfies
$\bar C(p^\star)<B$. By local outcome growth, a policy $p'$ arbitrarily
close to $p^\star$ has strictly higher welfare. Taking
$\lVert p'-p^\star\rVert_\infty<B-\bar C(p^\star)$ keeps $p'$ within
budget, a contradiction. Thus every budget-constrained maximizer has cost
exactly $B$.
Any such maximizer attains $T^\star$ at cost $B$, so $\bar C^\star(T^\star)\le B$.
If $\bar C^\star(T^\star)<B$, there is a policy $q$ with $\bar C(q)<B$ and
$\bar Y(q)\ge T^\star$. Such a policy is also a budget-constrained
maximizer but does not exhaust the budget, contradicting the preceding
argument. Hence $\bar C^\star(T^\star)=B$. A policy optimal for either formulation
is feasible and optimal for the other, proving equality of the optimizer sets.
\end{proof}

\subsection{Proof of Theorem \ref{thm:dual_regret}}
\begin{proof}
For part (i), write $V_\epsilon^{W,\star}=\sup_p V^W_\epsilon(p)$.
Because $h^+_\epsilon(c;B)\ge\bm{1}\{c>B\}$,
$V^W_\epsilon(p)\le R^W(p)$ for all $p$. Further, every policy with
$\bar C(p)\le B-\epsilon$ has zero hinge penalty, so
$V_\epsilon^{W,\star}\ge \bar Y^\star(B-\epsilon)$. Empirical optimality
and the $2/\epsilon$-Lipschitz property of the hinge imply
\begin{align*}
 \bar Y^\star(B)-R^W(\widehat p^W_\epsilon)
 &\le \bar Y^\star(B)-V_\epsilon^{W,\star}
 +2\sup_{p\in\mathcal P}
     |\widehat V^W_\epsilon(p)-V^W_\epsilon(p)|\\
 &\le \bar Y^\star(B)-\bar Y^\star(B-\epsilon)
 +2\sup_p|\widehat{\bar Y}^{\ipw}(p)-\bar Y(p)|
 +\frac{4C_y}{\epsilon}
       \sup_p|\widehat{\bar C}(p)-\bar C(p)|.
\end{align*}
Taking expectations and applying the expected uniform deviations in
Lemmas 1 and 2 proves \eqref{eq:dual_general_regret}.

For part (ii), let $E_C$ and $E_Y$ be the uniform-deviation events
$\sup_p|\widehat{\bar C}(p)-\bar C(p)|\le b_C(\delta)$ and
$\sup_p|\widehat{\bar Y}^{\ipw}(p)-\bar Y(p)|\le b_Y(\delta)$.
By Lemmas C.1 and C.2 and a union bound,
$\Pr(E_C\cap E_Y)\ge1-\delta$. Fix any
$q\in\mathcal P$ with
$\bar C(q)\le B-\epsilon-b_C(\delta)$.
On $E_C$, $\widehat{\bar C}(q)\le B-\epsilon$, so its
empirical hinge penalty is zero. Empirical optimality therefore gives
\[
 C_y h_\epsilon^+\{\widehat{\bar C}(\widehat p^W_\epsilon);B\}
 \le \widehat{\bar Y}^{\ipw}(\widehat p^W_\epsilon)
      -\widehat{\bar Y}^{\ipw}(q).
\]
On $E_Y$, the right-hand side is bounded above by
$2M+2b_Y(\delta)=C_y\alpha_N(\delta)$.
By the definition of the reflected hinge, the inequality
$h_\epsilon^+(c;B)\le\alpha<1$ implies
$c\le B-\epsilon k(\alpha)$.
Consequently,
\[
 \widehat{\bar C}(\widehat p^W_\epsilon)
 \le B-\epsilon k\{\alpha_N(\delta)\}
 \le B-b_C(\delta),
\]
so $\bar C(\widehat p^W_\epsilon)\le B$ on $E_C\cap E_Y$.
The same empirical optimality inequality, and the nonnegativity of the
penalty, imply
$\widehat{\bar Y}^{\ipw}(\widehat p^W_\epsilon)
 \ge\widehat{\bar Y}^{\ipw}(q)$.
Thus, on $E_C\cap E_Y$,
\[
 \bar Y(\widehat p^W_\epsilon)\ge\bar Y(q)-2b_Y(\delta).
\]
Taking the supremum over all such $q$ and using the population budget
feasibility just shown gives
\[
 \bar Y^\star(B)-R^W(\widehat p^W_\epsilon)
 \le \bar Y^\star(B)-\bar Y^\star\{B-\epsilon-b_C(\delta)\}+2b_Y(\delta).
\]
On the complement of $E_C\cap E_Y$, the regret is bounded above by
$2M+C_y$. Taking expectations yields \eqref{eq:dual_refined_regret}.
The rate follows because $b_C(\delta_N)$ and $b_Y(\delta_N)$ are
$O\{\sqrt{\log N/N}\}$, while
$k\{\alpha_N(\delta_N)\}$ converges to
$k(2M/C_y)>0$.
\end{proof}

\renewcommand{\refname}{References}
\putbib[appendix_bibliography]
\end{bibunit}
\end{document}